\documentclass[prb,reprint,aps]{revtex4-1} 

\usepackage{amsmath} 
\usepackage{amsfonts} 
\usepackage{graphicx}
\usepackage[english]{babel}
\usepackage{amsmath,amssymb}
\usepackage{gensymb}
\usepackage{xcolor}
\usepackage{natbib}
\usepackage{siunitx}

\begin{document}

\title{Risk assessment for long and short range airborne transmission of SARS-CoV-2, indoors and outdoors}
\date{\today}


\author{Florian Poydenot\textsuperscript{a}}
\author{Ismael Abdourahamane\textsuperscript{a}}
\author{Elsa Caplain\textsuperscript{a}}
\author{Samuel Der\textsuperscript{a}}
\author{Jacques Haiech\textsuperscript{b}}
\author{Antoine Jallon\textsuperscript{a}}
\author{In\'es Khoutami\textsuperscript{a}}
\author{Amir Loucif\textsuperscript{a}}
\author{Emil Marinov\textsuperscript{a}}
\author{Bruno Andreotti\textsuperscript{a}}
\email{andreotti@phys.ens.fr}
\affiliation{\textsuperscript{a}Laboratoire de Physique de l’École normale supérieure, ENS, Université PSL, CNRS, Sorbonne Université, Université Paris Cité, F-75005 Paris, France.\\
\textsuperscript{b}Cogitamus Laboratory and CNRS UMR 7242 BSC, 300 Bd S\'ebastien Brant, CS 10413, 67412 Illkirch Cedex.}

\begin{abstract}
Preventive measures to reduce infection are needed to combat the COVID-19 pandemic and prepare for a possible endemic phase. Current prophylactic vaccines are highly effective to prevent disease but lose their ability to reduce viral transmission as viral evolution leads to increasing immune escape. Long term proactive public health policies must therefore complement vaccination with available nonpharmaceutical interventions (NPI) aiming to reduce the viral transmission risk in public spaces. Here, we revisit the quantitative assessment of airborne transmission risk, considering asymptotic limits that considerably simplify its expression. We show that the aerosol transmission risk is the product of three factors: a biological factor that depends on the viral strain, a hydrodynamical factor defined as the ratio of concentration in viral particles between inhaled and exhaled air, and a face mask filtering factor. The short range contribution to the risk, both present indoors and outdoors, is related to the turbulent dispersion of exhaled aerosols by air drafts and by convection (indoor), or by the wind (outdoors). We show experimentally that airborne droplets and CO$_2$ molecules present the same dispersion. As a consequence, the dilution factor, and therefore the risk, can be measured quantitatively using the CO$_2$ concentration, regardless of the room volume, the flow rate of fresh air and the occupancy. We show that the dispersion cone leads to a concentration in viral particles, and therefore a short range transmission risk, inversely proportional to the squared distance to an infected person and to the flow velocity. The aerosolization criterion derived as an intermediate result, which compares the Stokes relaxation time to the Lagrangian time scale, may find application for a broad class of aerosol-borne pathogens and pollutants.
\end{abstract}


\maketitle

\textit{Making rational public health policy decisions to prevent the dominant routes of viral transmission requires a simple but quantitative assessment of the airborne transmission risk in public places, both indoors and outdoors. Here we show that CO$_2$ and aerosol viral particles disperse following the same law, which allows us to relate this aerosol transmission risk to the CO$_2$ concentration. The results provide quantitative guidance useful to complement vaccination, treatment for the vulnerable patient population and the test-trace-isolate strategy by a national ventilation plan and gradual mandates of FFP2 respirators in indoor places, when the epidemic circulates above a threshold.}

\section{Introduction}
The SARS-CoV-2 pandemic enters its third year despite the design of highly effective vaccines, which induce circulating antibodies and systemic T and B cell responses that block viral spread and disease. Beside the lack of vaccination at the global scale, they do not establish immunity at mucosal surfaces against infection by variants like Omicron (B.1.1.529), which present both increased intrinsic transmissibility and viral immune evasion. Widespread transmission therefore contributes to a degree of unpredictability in the evolution of the pandemic. As a consequence, vaccination must be complemented in the long term by effective public health policies contributing to suppress transmission at a low economic and social cost. To help making rational public health policy decisions, we propose here a method to measure the risk for long- and short-range airborne transmission of SARS-CoV-2, both indoors and outdoors.

Respiratory viruses and bacteria can be transported by droplets emitted by coughing or sneezing, which may cause symptomatic transmission, and during expiratory human activities such as breathing, speaking or laughing, which may cause asymptomatic and pre-symptomatic transmission. Pathogens responsible for illnesses such as influenza, tuberculosis, measles or SARS, initially carried by these droplets, can form an aerosol phase \cite{zhou_defining_2018,fennelly_particle_2020} and cause airborne transmission by inhalation. The silent spread of SARS-CoV-2 by asymptomatic infected individuals, who do not cough nor sneeze, has been hypothesized as early as January 2020. Since June 2020, there is ample evidence that this virus is primarily transmitted through aerosols \cite{morawska_it_2020,zhang_identifying_2020,greenhalgh_ten_2021,jimenez_what_2022}. This is particularly obvious in public places where face mask wearing has been mandatory \cite{cheng_face_2021} as the heavier, millimeter-sized droplets have a ballistic trajectory that is relatively insensitive to the presence of air and are stopped by all types of masks.

Many misconceptions regarding aerosols can be found in the medical and scientific literature. Clinicians often use an incorrect definition, still reported by the World Health Organization: aerosols would be particles smaller than $5 \; {\rm \mu m}$ that settle slowly enough to be transported over a few meters \cite{jimenez_what_2022,randall_how_2021}. As a consequence, problematic criteria are introduced, such as the distance after which a particle launched with an initial velocity in a fluid at rest stops \cite{nazaroff_indoor_2022}, or the settling time of a single particle dropped from head height in still air \cite{merhi_assessing_2022}. An aerosol is a locally homogeneous phase constituted of solid or liquid particles suspended in a gas, either by thermal fluctuations (for very small particles) or by turbulent fluctuations. The aerosol phase tend to homogenize and to diffuse over the whole available space by turbulent dispersion. The "slow settling" misconception therefore omits the fundamental constitutive mechanism of aerosols of any particle size: turbulence.

The infection risk has been modeled in a series of papers, starting from the seminal works of Wells and Riley \cite{wells_airborne_1955,riley_airborne_1978}, which parametrize the infection risk as a function of the global air dilution and the disease infectiousness. The authors work upon the the well-mixed hypothesis, in which air inside a room is instantly mixed by turbulence so that all people inside breathe the same air. Beyond this hypothesis, fluid dynamics models of indoor air circulation \cite{ho_modeling_2021,li_probable_2021,vuorinen_modelling_2020,liu_simulation-based_2021,wang_short-range_2021} and heterogeneous airflows produced by respiratory activity have been introduced \cite{bourouiba_fluid_2021,yang_towards_2020,giri_colliding_2022}. The Wells-Riley model has been extended to unsteady conditions by Rudnick and Milton \cite{rudnick_risk_2003}, who have pointed out that CO\textsubscript{2} could be used as a risk proxy \cite{bivolarova_comparison_2017}. Regarding SARS-CoV-2, most models extend the Wells-Riley equation to account for unsteady conditions in ventilation or occupancy \cite{bazant_monitoring_2021,burridge_predictive_2021} and to different viral emission rates depending on the respiratory activity \cite{bazant_guideline_2021}, face coverings and particle removal and inactivation \cite{peng_exhaled_2021,peng_practical_2022,buonanno_estimation_2020}. All of these models use well-characterized ``super-spreading'' events to derive estimates of the viral emission rate, or closed micro-societies like cruise ships \cite{azimi_mechanistic_2021,parhizkar_quantitative_2021}.

These models have left three problems open up to now. (i) Can an analytical closed formula be derived, simple enough to be used to produce regulatory ventilation standards? (ii) Is the dispersion of CO\textsubscript{2} and airborne viral particles governed by the same law? (iii) What is the law governing the short range contribution to the transmission, both indoor and outdoor? Here, we propose a definition of the environmental risk of viral transmission in public spaces such as schools, offices, university lecture halls, museums or shopping centers, but also outdoor. We report experimental and theoretical results showing that exhaled CO\textsubscript{2} and airborne viral particles diffuse at the same rate. We finally derive a quantitative analytical model which relates this risk to the CO\textsubscript{2} concentration.
\begin{figure}[t!]
\includegraphics{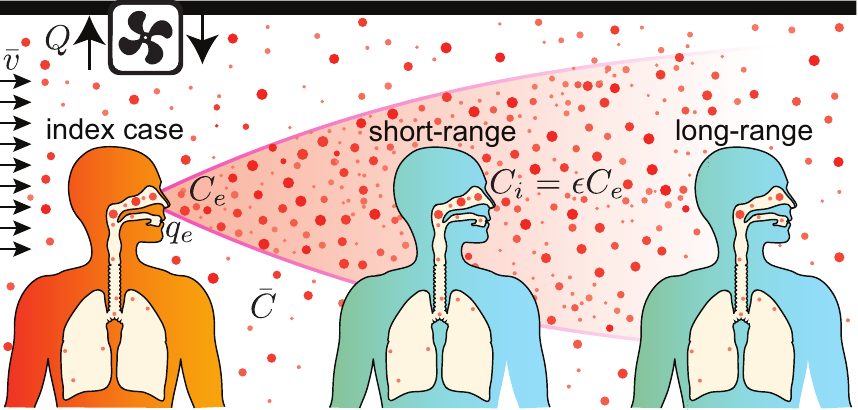}
\caption{Viral particles are exhaled by an infected person at a concentration $C_e$. They are dispersed by drafts (velocity $\bar v$), which lead to a decrease of their concentration. The dispersion cone must not be confused with the conical shape of high speed jets in a fluid at rest \cite{abkarian_stretching_2020,abkarian_speech_2020,yang_towards_2020}. In an indoor space, viral particles are stored, which translates into a background homogeneous concentration $\bar C$, controlled by the ventilation rate $Q$.}
\label{fig:Schematic}
\end{figure}
\section{Airborne transmission mechanism}
The main entry of SARS-CoV-2 virus is through the upper respiratory epithelium. To colonize a cell, an inhaled viral particle, interacts through the spike protein -- which is cleaved by a host cell protease, mostly the TMPRSS2 protease for the wild Wuhan-1 strain -- with a host cell membrane protein, the ACE2 receptor. Cleavage of the spike protein is necessary for a conformational change so that it can effectively interact with the ACE2 receptor. This interaction leads to the formation of a virus-ACE2 complex which triggers the internalization of the virus inside the cell. It replicates its RNA molecule, produce the proteins required for self-assembly of new viral particles, which are released, leading to the colonization of neighboring cells \cite{snijder_unifying_2020}.

An organism may become infected if a sufficient amount of viral particles interact with cells expressing both the TMPRSS2 protease and the ACE2 receptor and if the virus is able to hack into cellular mechanisms to produce and disseminate new virions. From the upper respiratory epithelium, which is the first tissue to be infected (nasal epithelium for viral strains before Omicron, but also throat epithelium for Omicron, due to a weaker dependency to the TMPRSS2 protease), the virus, embedded in mucus, is carried to the trachea, to the lungs or the esophagus, and finally to deeper organs. It has also the possibility to reach the brain and to colonize certain cells of the cortex.

When the virus is concentrated in the nasal cavity, it is disseminated via a mist of fine droplets of mucus or saliva dispersed by breathing, talking or singing. A sneeze or cough produces larger droplets containing viral particles (Figure S3). The evaporation of mucosalivary liquid droplets in the air is controlled by the ambient relative humidity $\mathrm{RH}$ and by its content in surfactants, proteins and electrolytes. Initially, transport of water molecules from the droplets to the surrounding air is diffusive so that the drop squared radius decreases linearly in time. After a very short time, the droplet stabilizes at an equilibrium radius at which the viral particle is surrounded by proteins and water \cite{mikhailov_interaction_2004,pohlker_respiratory_2021,seyfert_stability_2022,chong_extended_2021,merhi_assessing_2022}. The influence of droplet chemical composition on the equilibrium radius is still poorly understood \cite{huynh_evidence_2022,vejerano_physico-chemical_2018,oswin_dynamics_2022-1}. The radius shrinks by a factor $2$ to $5$ (see Supplementary Information): droplets emitted during breathing, below $20\;{\rm \mu m}$, are therefore sub-micronic or micronic and stay suspended in the air.

The transmission risk increases with the intake viral dose $d$, defined as the amount of infectious viral particles inhaled by a person, cumulated over time. $d$ increases with the time of exposure to the virus, with the inhalation rate $q_e$, i.e. the product of the breathing rate by the tidal volume (for light exercise, $q_e\simeq 0.5\;{\rm m^3/hour}$), and with the concentration $C_i$ of infectious viral particles in the inhaled air. As a dose is a quantity of virus, it can be measured using quantitative RT-PCR and is then expressed in genome units (GU). However, it is better adapted to measure a dose by infecting a culture cell monolayer, in plaque-forming units (PFU). Plaque-forming units measure the ability of viral particles to replicate and to be secreted by the chosen cell type, while qRT-PCR measures the number of RNA molecules and is not sensitive to replication potential. On generic Vero cells, the amount of virus needed on average to form one cell lysis has decreased from $1400~{\rm GU/PFU}$ for the wild strain to $240~{\rm GU/PFU}$ for the Delta variant \cite{poydenot_crossroads_2022}.

The inhaled dose $d$ is the product of two factors:
\begin{itemize}
\item a purely biological factor reflecting the exhalation flux of viral particles and the ability of the virus to infect a person
\item a purely physical factor $\epsilon$ reflecting the dilution of viral particles between exhalation and inhalation. 
\end{itemize}
We will investigate these two factors independently in the next sections.

\section{Dispersion and transport of CO\textsubscript{2} and viral particles}
{\it Average concentration~--~}In a public space where many people are gathered, let us single out two people: one is infectious and exhales viral particles and the second one inhales the air. 
The air exhaled by the infectious person is concentrated in viral particles which are gradually diluted in the ambient air, in a way similar to the smoke of a cigarette. Outdoors, far from an infected person, the viral particle concentration vanishes. However, indoors, viral particles accumulate: the concentration in the wake of an infected person decays from its exhalation concentration $C_e$, which is high, to a constant concentration $\bar C$ far away. The dilution factor $\epsilon$ is defined as the ratio of the viral concentration $C_i$ in the inhaled air and the viral concentration in exhaled air $C_e$ (Figure~\ref{fig:Schematic}). Consider first a closed room of volume $V$ where there is a single person exhaling viral particles at a concentration $C_e$. A ventilation of flow rate $Q$ replaces exhaust air at the average concentration $\bar C$ by fresh air. The average concentration $\bar C$ obeys the conservation equation: $V \mathrm d\bar C/{\mathrm dt}= q_e C_e- Q \bar C$. It is a linear relaxation equation whose exponential relaxation time is $V/Q$. The average dilution factor tends towards the steady state solution $\bar \epsilon= \bar C/C_e=q_e/Q$ \cite{rudnick_risk_2003}.

\begin{figure}[t!]
\centering
\includegraphics{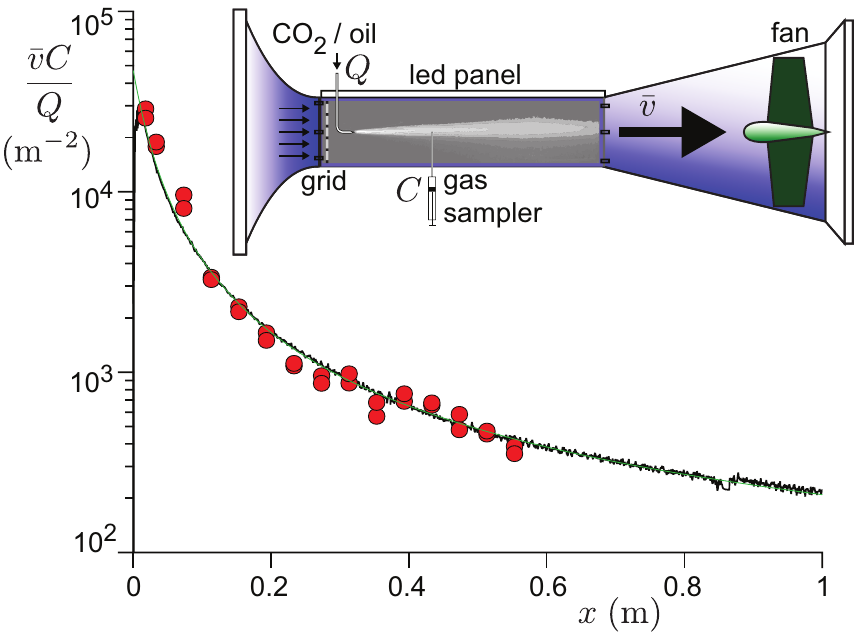}
\caption{Rescaled concentration profile of $10\;{\rm \mu m}$ size oil droplets (black solid line) and CO\textsubscript{2} (Red circles) in a wind tunnel experiment. The CO\textsubscript{2} concentration along the axis is measured after sampling the gaz with a syringe. The oil droplet concentration is measured using the optical absorption of light. CO\textsubscript{2} and oil aerosol are introduced successively at a rate $Q$, $20~{\rm cm}$ downstream of the turbulence generating grid. The working velocity is $\bar v=9\;{\rm m/s}$. The green line is the best fit by Eqs~(\ref{eq:ConcentrationGauss}-\ref{eq:EqDiffusionRadius}).}
\label{fig:CO2Oil}
\end{figure}

{\it Dispersion~--~}We have investigated experimentally and theoretically the turbulent dispersion of viral particles and CO\textsubscript{2} in a turbulent flow of mean velocity $\bar v$, characterized by a root mean square velocity $\sigma_V$. This corresponds to a generic situation where horizontal drafts (indoors) or wind (outdoors) dominate over thermal plumes, so that natural convection can be neglected. We have for instance shown in a recent paper that this was the case in the corridors of commercial malls \cite{poydenot_turbulent_2021-1}. The dispersion of CO\textsubscript{2} and of oil droplets of typical size $10~{\rm \mu m}$ is studied separately, but in the same conditions, in the wind tunnel schematized in Figure~\ref{fig:CO2Oil} (b). The smoke concentration field is measured using high resolution pictures ($7380 \times 4920$) with a $2\;{\rm s}$ exposure time (Figure S1). The CO\textsubscript{2} concentration profile $C^{{\rm CO}_2}(x)$ along the axis is measured after careful sampling of air using a syringe: the air is transferred into a chamber where a vacuum has been set, equipped with a CO\textsubscript{2} non-dispersive infrared sensor (Figure S2).

As dispersion is slow compared to the axial convection, the axial coordinate $x$ is equivalent to time $t=x/\bar v$. Transverse diffusion is controlled by random motion and leads, according to the central limit theorem, to a quasi-Gaussian radial profile of concentration \citep{poydenot_turbulent_2021-1}. As the flux across a transverse section is equal to the source emission rate $q_e C_e$, the concentration field takes the form:
\begin{equation}
C=C_e\;\frac{q_e }{\pi \sigma_R^2 \bar v}\;\exp\left(-\frac{r^2}{2\sigma_R^2}\right)+\bar C
\label{eq:ConcentrationGauss}
\end{equation}
The measurement of the dispersion radius $\sigma_R$ as a function of $x$ gives the law of spatial decay of the concentration.

Because of turbulence, the velocity of an elementary volume of fluid is correlated along its Lagrangian trajectory. This drives transported particle velocities to be correlated too. We model the dispersion as a Langevin process characterized by an exponentially decaying Lagrangian correlation function: $\overline{\mathbf{v^{\prime}}(t)\mathbf{v^{\prime}}(t+\tau)}=\sigma_V^2\exp(-\tau/{\mathcal T})$, where ${\mathcal T}$ is the Lagrangian integral time scale. The dispersion of fluid particles injected at a source point at time $t=0$ is given by Taylor's theorem, which gives at time $t=x/\overline v$ \cite{poydenot_turbulent_2021-1}:
\begin{equation}
\sigma_R^{2}= \frac{2}{3}\sigma_V^2 {\mathcal T}^2 \left[\exp \left(-\frac{x}{{\overline v}{\mathcal T}}\right) +\frac{x}{{\overline v}{\mathcal T}}-1\right]
\label{eq:EqDiffusionRadius}
\end{equation}
At large distances $x \gg \overline v \mathcal T$, it predicts a diffusive regime $\sigma_R \sim \sigma_V \left( {\mathcal T} x/ {\overline v}\right)^{1/2}$, the concentration along the axis ($r=0$) decaying as $1/x$. The most important dispersion takes place at short distances $x \ll \overline v \mathcal T$. Transport obeys a ballistic-like regime of the form $\sigma_R\sim \sigma_V x/ \bar v$, which corresponds to a fast decay of the concentration along the axis ($r=0$) as:
\begin{equation}
C-\bar C=C_e\;\frac{q_e \bar v}{\pi (\sigma_V x)^2}.
\label{eq:EqUnssquare}
\end{equation}

{\it Face masks and respirators~--~}In public places where face masks are mandatory, the aerodynamical contribution to $\epsilon$ must be multiplied by the product of the exhalation and inhalation mask filtration factors. All types of masks totally filter droplets of a fraction of millimeter. However, they have very different efficiencies between $0.1\;{\rm \mu m}$ and $0.5\;{\rm \mu m}$, which is the range of equilibrium sizes of the smallest mucus droplets after evaporation. Mask efficiency presents a minimum around $0.3\;{\rm \mu m}$, where the particles are large enough to make Brownian collection inefficient, and tiny enough to have small inertia, insufficient to make them leave the streamlines wrapping around the fibers \cite{hinds_aerosol_1999}. Filtration factors are determined by the material properties and the respiratory activity at play, but most importantly by proper mask fit. Cloth and surgical masks tend to be much looser fitted than respirators, thereby greatly reducing their filtration factor compared to what their fabric could achieve alone \cite{cappa_expiratory_2021}. Cloth masks have an effective filtration factor around $\lambda = 0.70$, on average \cite{pan_inward_2021,hill_testing_2020}. Assuming they are well-fitted, surgical masks have an effective filtration factor $\lambda = 0.28$ \cite{lindsley_comparison_2021,oberg_surgical_2008} and N95/FFP2 respirators $\lambda = 0.10$ \cite{qian_performance_1998,asadi_efficacy_2020}.

\section{Modeling the transmission risk}
{\it Intake viral dose~--~}Transmission risk assessment requires the estimate of the probability of infection under a given intake viral dose $d$. The simplest hypothesis is to assume an independent action of all inhaled replicable viral particles, which means that a single virus can initiate the infection. However, more than one is statistically needed, as the probability that a single infectious viral particle overwhelms the host immunity defences successfully is small, typically between $10^{-3}$ and $10^{-2}$ \cite{poydenot_crossroads_2022}. For a person having inhaled an intake dose $d$, the probability law of infection $p(d)$ takes the form:
\begin{equation}
p(d)=1-e^{-a d}
\end{equation}
where the susceptibility $a$ is the inverse of the infection dose defined, for each individual, as the intake dose for which the probability of infection is $1-1/e\simeq 63\;\%$. $a$ is widely distributed across individuals, according to a probability distribution $f(a)$. We denote by $\bar a=\int a f(a) \mathrm{d}a$ the population average of $a$. $\bar a^{-1}$ is the infectious dose, called the quantum of infection when used as a convenient unit for a quantity of infectious viral particles; the product $\bar a d$ is then the dose expressed in ``quanta''. For the wild strain Wuhan-1, the infectious quantum has been estimated around $5.6 \times 10^5 {\rm GU}$ and $400\;{\rm PFU}$ \cite{poydenot_crossroads_2022}. Importantly, the distribution $f(a)$ includes the effect of vaccinal or infection induced immunisation. The infectious quantum is indeed a population average over different immunity conditions.

{\it Relation between inhaled dose and viral exhalation rate~--~}In the literature \cite{peng_exhaled_2021,vouriot_seasonal_2021,kriegel_predicted_2020,lelieveld_model_2020,bazant_guideline_2021,pohlker_respiratory_2021,vuorinen_modelling_2020} devoted to airborne transmission of SARS-CoV-2, most authors have considered that the viral emission rate, i.e. the number of virus exhaled per unit time, is the product of the volume of mucus droplets emitted per unit time by the viral load in the nasal cavity for mucus droplets, in the throat for saliva droplets \cite{buonanno_estimation_2020} or deeper in the lungs for respiratory aerosols. This excludes the possibility of a viral enrichment at the interface between mucosalivary fluid and air, which could lead to a viral content of droplets which is not proportional to their initial volume.

\begin{figure}[t!]
\centering
\includegraphics{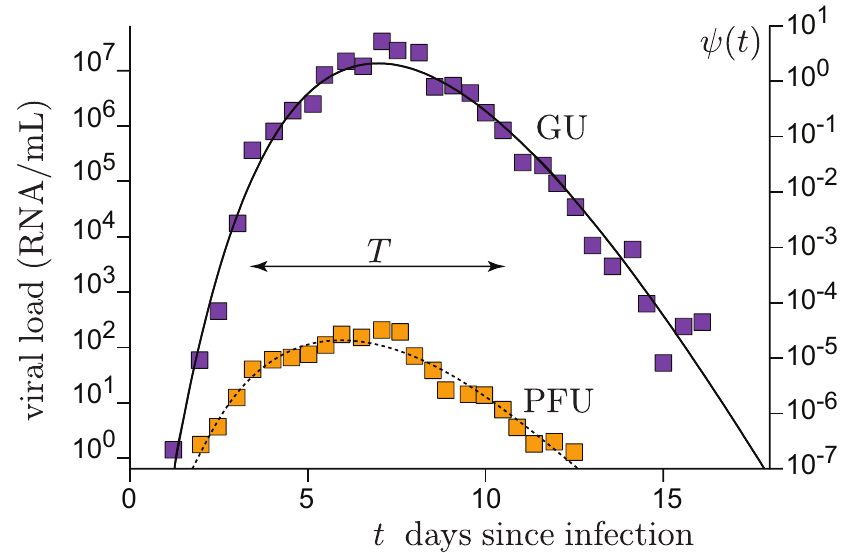}
\caption{Viral kinetics of SARS-CoV-2. Average viral load in the nasal cavity as a function of time since infection from Killingley, \textit{et al.} \cite{killingley_safety_2022}, using a strain close to Wuhan-1. Blue squares are obtained by quantitative RT-PCR and expressed in genom units (GU). Orange squares are obtained by counting lysis plaques and expressed plaque forming units (PFU). Right axis: Model dimensionless viral load $\psi(t)$ as a function of time $t$, in days, after infection. }
\label{fig:ViralLoad}
\end{figure}

Infectivity is proportional to the concentration of infectious viral particles in the exhaled air, noted $C_e$. It is assumed here to follow the same kinetics as the infectious viral load. It also depends on biological factors which differ from one viral strain to the other. The viral kinetics results from the competition between viral replication and immune response. For simplicity, $C_e$ can be assumed to present the same temporal profile amongst patients:
\begin{equation}
C_e=C_m\;\psi\left(t-t_c\right)
\end{equation}
where $C_m$ is a characteristic concentration and $t_c$ the infection time. Both $C_e$ and $C_m$ are expressed in ${\rm PFU/m^3}$. The rescaled viral load curve $\psi(t)$ is dimensionless and can be considered as the average over the sub-population considered. Typically, $\psi$ increases exponentially in the presymptomatic period and decreases exponentially at long times, as shown in Figure~\ref{fig:ViralLoad} \cite{jang_viral_2021,jones_estimating_2021,sethuraman_interpreting_2020}. We choose here the following normalization for $\psi(t)$, which provides an unambiguous definition of the mean contagious time $T$:
\begin{eqnarray}
\int_0^\infty \psi(t)\mathrm{d}t=T, \quad \int_0^\infty t\;\psi(t)\mathrm{d}t=T^2
\end{eqnarray}
The characteristic viral concentration $C_m$ is highly variable among infected people. Moreover, it depends on the vaccination status, on former infections, on the quality of the immune system, etc. In order to consider an average over all statistical realizations, we introduce the probability density function $g(C_m)$, which is typically log-normal. Depending on the hypotheses used to extract $C_m$ from sparse RT-qPCR measurements, its dispersion is found between $0.1$ to $1$ $\log_{10}\;{\rm copies/mL}$ above and below average \cite{jones_estimating_2021,monel_release_2021,elie_inferring_2021}.

The inhaled dose can be expressed as:
\begin{equation}
d= \int dt\;\lambda^2\;q\;C_i=\int dt\;\lambda^2\;\epsilon \; q \;C_m
\label{eq:defdose}
\end{equation}
The key quantity is the mean integrated viral emission $\bar h$, which accounts for all the biological part of the risk. By definition, it is the dose in ``quanta'' exhaled by a patient during the entire infectious period:
\begin{equation}
\bar h=\int dt \bar q \bar a C_e=\bar q \bar a \bar C_m T
\end{equation}
It is defined as an average over the sub-population attending the public space considered and may depend on the particular activity taking place in the public space through the mean inhalation rate $\bar q$. The integrated viral emission $\bar h$ is between $450$ and $500$ quanta for Wuhan-1, $800$ for Alpha, $1500$ for Delta, $2800$ for Omicron BA.1 and $4700$ for Omicron BA.2 \cite{poydenot_crossroads_2022}. 

 {\it Environmental risk~--~}We define the transmission risk of a public space as the average number of infections $r$ that an infected person staying in for an infinitely long time would cause, on average. This public space contributes to epidemic decay if $r < 1$ and to its growth otherwise. $r$ characterizes the environmental conditions and behaviors in the public space. We consider a public place hosting $N$ people during a given period of time, among which $M$ are infectious. The epidemic incidence $I(t)$ is the fraction of a given population infected per unit time. In practice, the fraction of the population $I(t) \mathrm \delta t$ infected between $t$ and $\delta t$ is measured with $\delta t=1\;{\rm day}$, although conceptually $\delta t$ is infinitesimal. The prevalence $P$ is the fraction of this population currently infectious and is related to $I$ by the equation:
\begin{equation}
P(t)=\int_{-\infty}^0 I(t') \psi(t-t')\; \mathrm d t'
 \end{equation}
On average, the number $M$ of infectious people among $N$ is $\bar M=PN$. Under the Poissonian hypothesis, the mean number $rM$ of people infected reads
\begin{equation}
rM=\sum_{i=1}^{N-M}\left(1-e^{-a_{i} d_{i}}\right)
\end{equation}
where $d_i$ is the intake dose of the individual labelled $i$ while $1/a_i$ is their infection dose. In the case where all $N$ people are statistically subjected to the same intake dose $d_i=d$, the risk $r$ reads:
\begin{equation}
r=(N-M) \int_0^\infty f(a) \left(1-e^{-a d}\right)\; \mathrm d a
\label{EqInte}
\end{equation}
The quantity $\rho(d)=\int f(a) \left(1-e^{-a d}\right) \; \mathrm{d} a$ is the probability to get infected when an intake dose $d$ is inhaled and is called the dose response function. We consider now the limit where the intake dose, expressed in quanta $\bar a d$, has a very low probability of being larger than $1$. This excludes super-spreading events, which occur when one infectious person (or several) with a large exhaled concentration $C_m$ attends an under-ventilated place, leading to multiple simultaneous infections. Then, performing the linearization $1-\exp(-ad) \simeq ad$, the equation simplifies into $\bar Z=(N-M) \bar a d=M(N-M) \bar a \epsilon \bar q \bar C_m T$. The average number of secondary infections $r=\bar Z/M$ is therefore proportional to the biological factor $\bar h$, to the dilution ratio $\epsilon$:
\begin{equation}
\bar r=(N-M)\epsilon \bar h
\label{EqLinea}
\end{equation}
For the same concentration of viral particles in the air, the transmission risk increases with the number of people $(N-M)$ susceptible to be infected as they all have an equal individual risk of being infected.

\section{Results}

{\it Transport of respiratory aerosols~--~}Does the dispersion of CO\textsubscript{2} and virus laden droplets obey the same law? To answer this question, we compare experimentally and theoretically the turbulent dispersion of an aerosol of oil droplets and of CO\textsubscript{2} in a controlled turbulent flow. Figure~\ref{fig:CO2Oil} (a) shows that the concentration decays of the gaz and of the aerosol superimpose without any adjustable parameter. This remarkable result is highly non-trivial. Inertial particles disperse at the same rate as massless particles if the particle velocity correlation is close enough to the fluid velocity correlation function. Inertia acts as a low-pass filter of the fluid velocity, with a cut-off time controlled by the Stokes times $\tau_S = \rho_p d^2/18 \eta$, which is the particle response time to a change in the fluid velocity \cite{bec_turbulent_2010}. 

The quantitative criterion for the aerosolization of particles of different sizes size must compare the particle inertia to the dispersion by turbulent fluctuations. The stopping distance after which a droplet exiting the upper airways with a large initial velocity stops in a fluid at rest \cite{hinds_aerosol_1999} includes inertia but ignores turbulent particle diffusion. Inertia acts all along the particle trajectory: the effect of particle size on turbulent dispersion is to make large enough particles slow-paced at changing their velocities, so that they leave the gas streamlines and decorrelate from the flow. We have shown here that the relevant turbulent property to which inertia must be compared is the Lagrangian time-scale $\mathcal T$. This experimental result is in agreement with correlations functions found from direct numerical simulations of particles in turbulence \cite{biferale_lagrangian_2008}. Particles form an aerosol phase whenever turbulence can homogenize their concentration. Settling gradually depletes the phase, but does not determine whether droplets should be regarded as individual ballistic particles or as an Eulerian phase. Settling times incorrectly assume that the air is still. Our experiments show that gravity is not the relevant physical mechanism that separates small particles from large particles, but the balance of inertia and turbulence is, and the dimensionless ratio $\mathrm{St} \equiv \tau_S/\mathcal T$ is the correct way of defining the Stokes number for the dispersion problem (see Supplementary Information for a derivation). If the Stokes number $\mathrm{St}$ is much smaller than $1$, particles presents a negligible inertia, and are therefore aerosolized and dispersed according to Equation~(\ref{eq:EqDiffusionRadius}).

In the wind tunnel used in our experiments, we measured $\mathcal T = 9\;{\rm ms}$ \cite{poydenot_turbulent_2021-1}. The crossover above which inertia becomes important ($\tau_S = \mathcal T$) is therefore expected for a droplet diameter $\sim 50\;{\rm \mu m}$ much larger than the oil droplet diameter. For a meter scale ventilated room, the Lagrangian timescale $\mathcal T$ is in the range of $10^2\;{\rm ms}$ and the cross-over droplet diameter is larger than $\sim 100\;{\rm \mu m}$. To conclude, CO\textsubscript{2} and viral particles are dispersed at the same rate so that the dimensionless factor $\epsilon$ can indeed be determined quantitatively from the measurement of the CO\textsubscript{2} concentration.

Figure~\ref{fig:CO2Oil} (a) shows that Equations~(\ref{eq:ConcentrationGauss}-\ref{eq:EqDiffusionRadius}), with a single relaxation time-scale $\mathcal T$, perfectly fit the data. This result is particularly subtle and surprising. At such high Reynold numbers (between $10^5$ and $10^6$), turbulence is fully developed in space. According to the energy cascade picture of turbulence, kinetic energy is injected at large scale and dissipated by viscosity at small scale. At intermediate length scales, called the inertial range, energy is transferred by inertial effects \cite{lesieur_turbulence_2008}, which cause non-trivial velocity correlations between two points separated in space. The typical flow evolution time scale at the dissipation scale is called the Kolmogorov time $\tau_K$ and is used in the turbulence community to define a different particle Stokes number $\mathrm{St}$ \cite{biferale_lagrangian_2008}. Indeed, $\tau_S/\tau_K$ is relevant for the pair dispersion of particles \cite{richardson_atmospheric_1926,bourgoin_turbulent_2015}. The transport of a single particle is a Lagrangian problem \cite{poydenot_turbulent_2021-1}, and therefore does not involve the Kolmogorov time, but rather the Lagrangian decorrelation time $\mathcal T$. Surprisingly, for the Reynolds numbers accessible numerically and experimentally, Lagrangian two-points correlation functions in time do not feature any inertial range, while Eulerian two-points space correlation functions do \cite{biferale_lagrangian_2008}. Figure~\ref{fig:CO2Oil} shows that we do not find any signature of an inertial range on the Lagrangian correlation function in time either. We do not observe any signature of a power-law regime, but instead, a simple ballistic-like regime. Simple arguments \`a la Kolmogorov would rather predict a much faster power-law decay of the velocity correlation function, leading quickly to a diffusive law and therefore of a decay of $C-\bar C$ as $x^{-1}$.

Experiments with a volunteer breathing through the mouth \cite{poydenot_turbulent_2021-1} have shown that viral particles are injected at a length scale $a \approx 0.3\;{\rm m}$, set by the head size, which is significantly smaller than the turbulent integral scale. Equation~(\ref{eq:EqUnssquare}) can directly be used for exhalations, by changing $x$ to $x+a$, so that:
\begin{equation}
\epsilon=\bar \epsilon +\frac{q_e \bar v}{\pi \sigma_V^2 (x+a)^2}
\label{eq:EqUnssquareEps}
\end{equation}
The aerosol concentration therefore present a fast decay as the inverse squared distance $x$ to the infected person. It is inversely proportional to the wind speed, at constant fluctuation rate $\sigma_V/\bar v \sim 10^{-1}$. Equation~\ref{eq:EqUnssquareEps} constitutes a central result of this paper, as it controls the spatial structure of the transmission risk outdoors.

{\it Transmission risk and CO\textsubscript{2} concentration~--~}The $N$ people present exhale CO\textsubscript{2} while only the $M$ infected people emit viral particles. As a consequence, the expression of $\epsilon$ in terms of the average CO\textsubscript{2} concentration $\langle C\rangle $ also contains a factor $N$:
\begin{equation}
\langle \epsilon\rangle =\frac{\langle C^{{\rm CO}_2}\rangle -C^{{\rm CO}_2}_\infty}{N\;C^{{\rm CO}_2}_e}
\label{eq:eqeps}
\end{equation}
where $C_e\simeq 37500\;{\rm ppm}$ is the average CO$_2$ concentration in exhaled air. It must be emphasized that Equation~(\ref{eq:eqeps}) does not assume any relationship between CO$_2$ and virus emission rates and is equally valid for breathing, talking or singing. It only hypothesizes that airborne viral particles are inhaled. The two factors balance each others in the transmission risk, which reads, after averaging over $M$: $\bar r=\lambda^2 (1-P)N \langle \epsilon\rangle \bar h$. The prevalence is in general much smaller than $1$ so that we obtain the final formula relating the viral transmission risk to the CO$_2$ concentration:
\begin{equation}
\boxed{\bar r= \lambda^2 \frac{\langle C^{{\rm CO}_2}\rangle -C^{{\rm CO}_2}_\infty}{C^{{\rm CO}_2}_e} \bar h}
\label{eq:rdec}
\end{equation}
This equation is remarkable: the dependence of the environmental risk, which characterizes the creation of transmission chains, with respect to the volume of the room, the ventilation or the occupancy number are all encoded in the space averaged CO\textsubscript{2} concentration $\langle C\rangle$. The result is more subtle than it seems at first sight. Indeed, the $N$ occupants of a public space all exhale CO\textsubscript{2} but only the $M$ infected ones exhale virions. Let us compare the risk of a well-ventilated lecture hall (say, at $750\; {\rm ppm}$ of CO\textsubscript{2}), with $50$ students, to the risk if the same students are spread out in two conventional rooms with the same CO\textsubscript{2} concentration. Obviously, all other things being equal, the probability of a student being infectious is twice lower when $25$ students are grouped together, rather than $50$. However, since the ventilation must be proportional to the room occupancy to keep the CO\textsubscript{2} level, the viral particles are twice more concentrated in the small room and therefore the inhaled dose doubles. Under the above assumption, the average number of people infected is the same, although the risk is distributed differently.

The average CO\textsubscript{2} concentration $\langle C^{{\rm CO}_2} \rangle$ can be defined to take into account both the long-range dilution $\bar \epsilon$ and the short-range contribution, present both indoor and outdoor, due to the higher aerosol concentration in the wake of infectious people. To achieve this, the CO\textsubscript{2} concentration must be averaged over the surface where one more person would stand in the public place. The reasoning is the same as before. Each time a CO\textsubscript{2} molecule reaches the measurement point, there is a probability $M/N$ that it has been exhaled by one of the $M$ infectious people. In that case, viral particles follow the same path as CO\textsubscript{2} molecules. The dilution between inhaled and exhaled viral particle concentrations is still valid, but only if averaged over the possible permutations of the $M$ infectious people amongst the $N$ people. Again, the transmission risk is distributed very differently in the wake of infectious and non-infected people but the average remains correct. The quantitative determination of the aerosol transmission risk is directly applicable to outdoor spaces, where it is limited to this dispersion cone in the wake of infected people. The short range transmission risk outdoors is generally smaller than indoors, due to larger air flow velocities.

Let us illustrate on two examples the consequences of equations (\ref{eq:EqUnssquare}) and (\ref{eq:EqUnssquareEps}). Consider a shopping mall corridor in which we have measured the air draft velocity to be typically $0.2\;{\rm m/s}$. This corresponds to an extra CO\textsubscript{2} concentration of $160\;{\rm ppm}$ at $1\;{\rm m}$, of $50\;{\rm ppm}$ at $2\;{\rm m}$, of $10\;{\rm ppm}$ at $5\;{\rm m}$. Consider now an outdoor situation with a $1\;{\rm m/s}$ wind. This corresponds to an extra CO\textsubscript{2} concentration of $30\;{\rm ppm}$ at $1\;{\rm m}$, of $10\;{\rm ppm}$ at $2\;{\rm m}$, of $2\;{\rm ppm}$ at $5\;{\rm m}$ (Figure S7).

\begin{figure}[t!]
\centering
\includegraphics{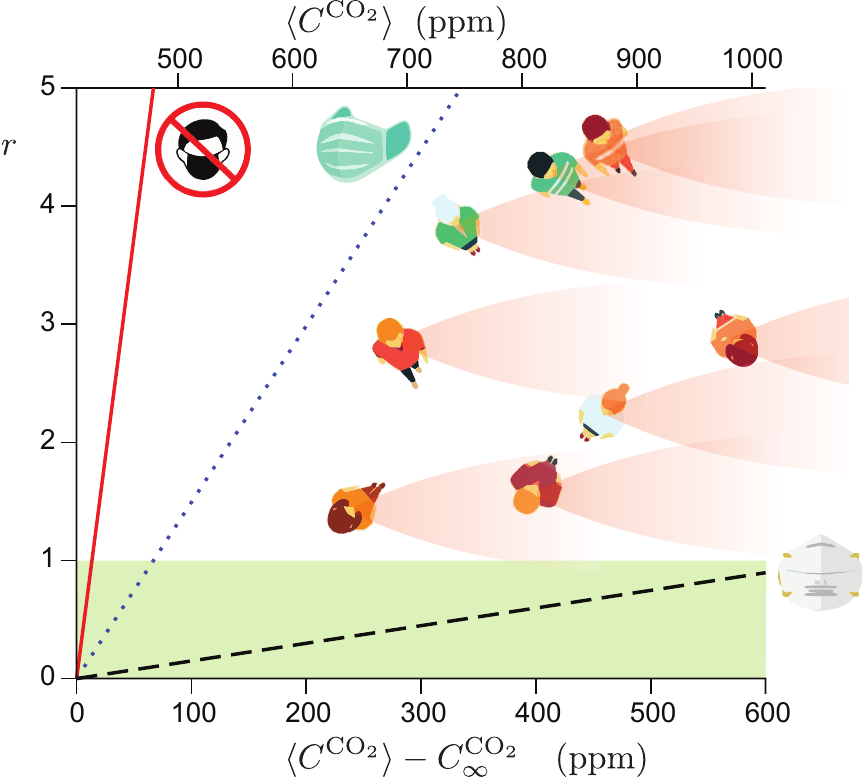}
\caption{Airborne transmission risk $r$ as a function of the concentration excess $\langle C^{\mathrm{CO}_2}\rangle - C^{\mathrm{CO}_2}_\infty$ for the Omicron BA.1 strain ($\bar h = 2800\;{\rm quanta}$). Solid line: no face mask. Dotted line: current level of mask-wearing in public, when mandatory ($\lambda^2=0.2$, Figure S4). Dashed line: FFP2 respirator ($\lambda^2=0.02$). Inset: schematic showing the spatial distribution of short-range transmission risk in a public space. $\langle C^{\mathrm{CO}_2}\rangle$ is by definition the space average of the CO\textsubscript{2} concentration over all places where a further person would spontaneously stand.}
\label{fig:risk}
\end{figure}

Figure~\ref{fig:risk} shows the linear relationship between the aerosol transmission risk and the CO\textsubscript{2} average concentration for different face masks and viral strains. The natural risk threshold is $r<1$, below which an infected person infects on average less than one other person, making the epidemic recede. With the current level of mask-wearing in public (Figure S4, Figure S5), which leads to a small filtration factor $\lambda^2=0.2$, the maximum excess CO\textsubscript{2} concentration should be $70\;{\rm ppm}$. This corresponds to a CO\textsubscript{2} concentration of at most $C=490\;{\rm ppm}$. If the risk is large but ventilation cannot be changed, N95/FFP2 respirators could be mandated. Social acceptability of such mandates could be managed by targeting first high-risk spaces, such as public transportation \cite{bertone_assessment_2022}, then gradually expanding the scope of non-pharmaceutical interventions based on reproduction number and prevalence thresholds, following transparent and planned rules. If respirators were mandated, the factor $\lambda^2$ would be decreased by $50$ and the maximum excess CO\textsubscript{2} concentration should be $670\;{\rm ppm}$. The risk becomes smaller than $1$ for standard ventilation conditions. Without mandatory face masks, the maximum excess CO\textsubscript{2} concentration should be $C=13\;{\rm ppm}$. This is far too small to be achievable in practice. Ventilation with fresh air is energy consuming, both in winter and during heat waves. It is therefore important to complement ventilation with HEPA filters or UV-C neon lights air purifiers (Figure S6), which are already used reliably and regularly \cite{raeiszadeh_critical_2020}, are cost-effective techniques for destroying nucleic acids, DNA or RNA from bacteria, viruses or other micro-organisms present in the air.

\section{Discussion}
We have obtained, using controlled approximations, a very simple expression showing a linear relationship between airborne transmission risk and CO\textsubscript{2} concentration. Several effects have been neglected, which we discuss here.

The effect of gravity is to transport droplets vertically at the settling velocity. It can be neglected when the latter is small in front of turbulent velocities. Typically, the $10\;{\rm \mu m}$ oil droplets settle at $3\;{\rm mm.s^{-1}}$. For the wind tunnel used here, the fluctuating velocity is typically $\sigma_V \sim 0.7~{\rm m.s^{-1}}$ which corresponds to a cross-over droplet diameter, above which settling is dominant, of $150~{\rm \mu m}$. For a meter scale ventilated room, the cross-over diameter remains of order $10^2~{\rm \mu m}$. For an aerosol of viral particles, deposition is therefore negligible, as can be checked from models taking it into account~\cite{bazant_guideline_2021,peng_exhaled_2021,bazant_monitoring_2021}. As evaporation is very fast, the effect of gravity is at worst transient and limited to the largest droplets.

We assume that dispersion is controlled only by the ambient air flow, and not by expiratory air flows. Unmasked coughs and sneezes can create large velocities that dominate over ambient transport, in which aerosols are exhaled in a buoyant jet that gets mixed at its boundaries \cite{bourouiba_fluid_2021,bourouiba_violent_2014}. Face masks greatly reduce expiratory velocities, so that transport by the ambient flow is recovered $20\;{\rm cm}$ away from the source \cite{deng_what_2021,verma_visualizing_2020,simha_universal_2020}. Our work is particularly applicable outdoor, where there is no mixing but rather a horizontal turbulent flow, and wind velocities easily dominate over exhalation (Figure S7). Measurements of crowds taking into account head orientation, crowd density and walking velocity show that horizontal velocities completely control the transmission risk \cite{garcia_model-based_2021,mendez_microscopic_2022}.

Virions transported in an aerosol phase are gradually degraded by the damage done by antiviral proteins in mucosalivary fluid \cite{malamud_antiviral_2011} and by UV-B radiation in sunlight or by UV-C. The inactivation rate of enveloped, airborne viruses such as SARS-CoV-2 depends on the physicochemical composition of the aerosol droplets, relative humidity $\mathrm{RH}$ and temperature in ways that are still unclear. Relative humidity plays an important role: the inactivation rate has a maximum at high $\mathrm{RH}$, reaches a minimum at intermediate $\mathrm{RH}$ then increases again for lower $\mathrm{RH}$ \cite{yang_dynamics_2011,tang_effect_2009,smither_experimental_2020,huynh_evidence_2022}. This suggests that virions can associate with proteins which protects them both from dessication and antivirals \cite{vejerano_physico-chemical_2018}. Ignoring viral inactivation is our main simplifying assumption outdoors, as sunlight may make it much faster. Inactivation can be modeled as a relaxation process of time-scale $\tau$, once $C$ is defined as the concentration of replicable viral particles: $\mathrm dC/{\mathrm dt}= (q_e C_e- Q \bar C)/V - \bar C/\tau$. The dilution factor $\epsilon$ and the transmission risk are therefore divided by $1+V/(Q \tau)$. The assumption made before corresponds to the regime where $Q \gg V/\tau$. As $\tau$ is on the order of an hour, this condition requires at least several ``air changes per hour'' (ACH) to be satisfied, which is the case in correctly ventilated public spaces. In steady state, measuring $\epsilon$ is therefore equivalent to measure both occupancy and ventilation rate. In transient conditions, often found in large public spaces, this approximation fails.

The mean integrated viral emission $\bar h$ has three sources of variability. The most important one is the variation of this biological factor between different public spaces, depending on the type of respiratory activity taking place there and on the characteristics of the population attending it. For breathing, the respiratory rate $q_e$ can vary by a factor $2$-$3$, depending on activity level and physiological characteristics. Here, it should be understood as the average value for the level of activity inside the public space, eg. its typical value at rest for a theater, light exercise for a shopping mall where people walk, and exercise for a sport facility.

The viral exhalation rate dependence on respiratory activity, talking and singing is scarcely known. It has been estimated that talking increases it by a factor $\sim 10$ \cite{coleman_viral_2022,shen_hybrid_2022}. However, few speech samples were culture-positive, either due to protocol limitations, or because the virus may have lost some of its ability to replicate when exiting through the mouth \cite{coleman_viral_2022,adenaiye_infectious_2021}. It is therefore difficult to define $\bar h$ for each public space, and it must be averaged over the whole society. Individual exhalation rates, peak viral load and individual infection doses are also broadly distributed, which creates another source of variability. There is also a variability originating from the number of contacts between individuals. Individuals who have many contacts get infected earlier and are thus removed from the pool of susceptible individuals, which decreases $\bar h$ over time.

The last simplifying assumption is a small intake dose $\bar a d$ (in quanta), which corresponds to a small dilution factor $\epsilon$. Equation~(\ref{eq:eqeps}) shows that this hypothesis corresponds to a situation which is either well ventilated or with a large number of people. This is generally the case for public spaces. By contrast, domestic spaces in winter generically gather few people in bad ventilation conditions, during long periods of time. This assumption excludes super-spreading events in which almost everyone gets infected because of these large deviations and poor ventilation.

\section{Conclusion}
 In this article, we have introduced an effective definition of the airborne transmission risk $r$ associated with a public space, defined as the average secondary infections per initially infected person. Under the commonly accepted hypothesis of no cooperation between virions, the risk is computed in the limit where it is low. It is related to the integrated quantum emission $\bar h$, to the mask filtration factor $\lambda^2$ and to the dilution factor $\epsilon$ between exhaled and inhaled air. $\bar h$ accounts for all biological aspects and depends on the viral strain considered but also on the characteristics of the sub-population attending the public space considered. In that sense, $\bar h$ takes into account the average respiratory activity in the public space in question. $\epsilon$ accounts for all hydrodynamical aspects and is decomposed into an average contribution controlled by ventilation and a spatially dependent contribution, localized in the dispersion cone of infected people, controlled by turbulent air flows. Indoor and outdoor spaces both present a risk of airborne transmission at short-range, in the dilution cone of the exhaled breath. But indoor, enclosed spaces only, present a risk of airborne transmission at long-range. Equation~(\ref{eq:EqUnssquareEps}) provides the answer to our initial question (iii) as it provides the law governing the short range contribution to the transmission, both indoor and outdoor.

Equation~(\ref{eq:rdec}) constitutes a central result of the article, as it provides an answer to our initial question (i). Indeed, it incorporates the ventilation flow rate, the room volume and the occupancy number into a single measurable quantity: the CO\textsubscript{2} concentration. The disappearance of the occupancy $N$ from Equation~(\ref{eq:rdec}) at large $N$ is non-trivial and comes from two factors balancing each other: on the one hand, CO\textsubscript{2} is exhaled by all individuals present, and not only by people infected by the virus; on the other hand, the transmission risk increases linearly with the number of people susceptible to be infected. As the risk is proportional to the CO\textsubscript{2} concentration, the acceptable CO\textsubscript{2} concentration below which $r<1$ can be readily adjusted as new variants with higher $\bar h$ keep on appearing, based on a feedback using the measured reproduction number.

The simplicity of Equation~(\ref{eq:rdec}) is based on the third important result derived here, both experimentally and theoretically: the dispersion of CO\textsubscript{2} and airborne viral particles are governed by the same law, for usual Reynolds numbers (question ii). The quantitative criterion is a small enough Stokes number $\mathrm{St} \equiv \tau_S/\mathcal T$ defined as the ratio of the Stokes time to the Lagrangian integral time. This result, albeit simple in appearance, results from a subtle effect of Lagrangian turbulence. This has important implications in determining infectivity times: as settling is not the relevant physical mechanism for aerosol transport, the infectivity of particles between $20\;{\rm \mu m}$ and $100\;{\rm \mu m}$ could be underestimated by models in which the air is still \cite{merhi_assessing_2022}. We are not aware of previous simultaneous measurements of gas and particles in the same controlled flow.

The transmission of SARS-CoV-2 in public spaces is predominantly airborne. In complement to vaccination, treatments for the vulnerable patient population and the test-trace-isolate strategy, the infection risk can be reduced by a combination of four collective actions:
 \begin{itemize}
 \item Ventilation with a sufficient fresh air flow rate per person to reduce the long-range risk.
 \item Monitoring CO\textsubscript{2} to measure the risk and adjust practices.
 \item Air purification to complement ventilation where needed.
 \item Turbulent dispersion, distancing and reduction of static crowds to reduce the short range risk, both indoors and outdoors.
 \end{itemize}
 
Ventilation guidelines or regulations in public spaces are centered around balancing thermal comfort with energy consumption and perceived air quality, not infection prevention \cite{morawska_paradigm_2021}, in part because the physical basis of airborne transmission was poorly understood \cite{randall_how_2021,jimenez_what_2022}. The approach presented here defines an unambiguous transmission risk, encoding most hydrodynamical aspects into a single measurable quantity. It is applicable to all airborne pathogens and to many flow configurations. It also defines a maximum acceptable risk: $r = 1$ is the threshold above which the epidemic continues to propagate. In that sense, it is a step towards ventilation regulation for infection prevention.

The transmission risk model outdoors is extremely easy to implement in practice. Combining Equations~(\ref{EqLinea}) and (\ref{eq:EqUnssquareEps}), we find
\begin{equation}
r=\frac{\bar h q_e \bar v}{\pi \sigma_V^2 (x+a)^2}.
\end{equation}
At the request of municipal services, we have applied it to the "Canal Saint-Martin" in Paris, France, the popular banks of a canal along which many people eat and drink. Measuring the mean distance between visitors (typically $1\;{\rm m}$), we find that when the wind is aligned with the banks, $r$ is below $1$ for the Wuhan-1 strain and for wind speeds above $1.5\;{\rm m/s}$. However, the wind is not always perfectly aligned, and fluctuations in the wind direction lead to dispersion over the canal, and therefore reduced risk. Recommendation to local policymakers was to emphasize the lower risk outdoors but simultaneously to mandate face masks during the very few days where there is not enough wind to disperse viral particles. Using a slightly revised dispersion formula, adapted to a cycling group, we have similarly been able to establish that the transmission risk is low in the Tour de France peloton, but not indoors, in the hotels hosting the cyclists.

We propose that the paradigm of airborne infection prevention policies could shift from a set of independent nonpharmaceutical interventions (mask mandates, occupancy limits, social distancing) to a standard of risk, translated to a maximum acceptable CO\textsubscript{2} concentration. Each public place has several options to meet this imposed standard, itself depending on the local community transmission, by lowering the maximum occupancy or by investing in better ventilation, by mandating masks or respirators, or by purifying the air with HEPA filters or UV-C flash lights (Figure S6). Restrictions can be gradually added at different reproduction rates and prevalence thresholds by targeting high-risk spaces first to balance social acceptability. Education on good quality masks and their wider availability could greatly reduce risk in public transportation \cite{bertone_assessment_2022} and shopping malls. If, after targeting first high-risk spaces, the global reproduction rate is still above $1$, restrictions can be expanded to lower-risk spaces, until the epidemic recedes. It would also be possible, for example, to mandate that mechanical ventilation in public buildings have two flow rates: one that allows for ordinary air renewal, keeping the CO\textsubscript{2} concentration below, say, $1000~{\rm ppm}$, and another much more powerful one designed for a drastic lowering of pathogen concentrations in the air, despite energy or comfort costs. Based on the results presented here, French shopping malls have successfully used smoke extractor fans, renewing the air in five minutes ($12\;{\rm ACH}$), to that effect since May 2021. Insofar as smoke extractors are compulsory in this type of public spaces, this has not resulted in any additional costs.


\section*{Acknowledgments}
The authors thank Joël Pothier and Alice Lebreton for the fruitful discussions.

\section*{Funding}
Unibail-Rodamco-Westfield on behalf of Conseil National des Centres Commerciaux (CNCC) has funded this work under the CNRS contract 217977, who asked the authors to make recommendations for a health protocol aiming to reduce and quantify the transmission risk in shopping centers. The conclusions of the present article are therefore of direct interest for the funding company. The authors declare no financial competing interest. The funding company had no such involvement in study design, in the collection, analysis, and interpretation of data, nor in the writing of the article. The authors had the full responsibility in the decision to submit it for publication.

\section*{Author contributions statement}
F.P., I.A., E.C., S.D., A.J., I.K., A.L., E.M. and B.A. conceived and conducted the experiments, all authors analyzed the results. B.A., J.H. and F.P. wrote and reviewed the manuscript.


\bibliography{ShortLongRangeCovid_nodoi}


\pagebreak

\onecolumngrid
\begin{center}
  \textbf{Risk assessment for long and short range airborne transmission of SARS-CoV-2, indoors and outdoors\\Supplementary Information}\\[.2cm]
  Florian Poydenot\textsuperscript{a}, Ismael Abdourahamane\textsuperscript{a}, Elsa Caplain\textsuperscript{a}, Samuel Der\textsuperscript{a}, Jacques Haiech\textsuperscript{b}, Antoine Jallon\textsuperscript{a}, Inés Khoutami\textsuperscript{a}, Amir Loucif\textsuperscript{a}, Emil Marinov\textsuperscript{a}, and Bruno Andreotti\textsuperscript{a*}
  {\itshape \textsuperscript{a}Laboratoire de Physique de l’École normale supérieure, ENS, Université PSL, CNRS, Sorbonne Université, Université Paris Cité, F-75005 Paris, France.\\
\textsuperscript{b}Cogitamus Laboratory and CNRS UMR 7242 BSC, 300 Bd S\'ebastien Brant, CS 10413, 67412 Illkirch Cedex.\\}
  ${}^*$Electronic address: \verb\andreotti@phys.ens.fr\\\
(Dated: \today)\\[1cm]
\end{center}
\twocolumngrid

\setcounter{equation}{0}
\setcounter{figure}{0}
\setcounter{table}{0}
\setcounter{page}{1}
\renewcommand{\theequation}{S\arabic{equation}}
\renewcommand{\thefigure}{S\arabic{figure}}
\renewcommand{\bibnumfmt}[1]{[S#1]}
\renewcommand{\citenumfont}[1]{S#1}

\section{Experimental methods}
\subsection{Wind tunnel}

Dispersion of aerosols and CO\textsubscript{2} is studied in a suction-flow wind tunnel with a $120 \times 23\times 23 \;{\rm cm}$ test section. A turbulence generating grid ($3\times 3$ grid of $7\times 7\;{\rm cm}$ squares separated by $1\;{\rm cm}$ wide bars) is place after the air inlet contraction section, $30\;{\rm cm}$ upstream of the test section. Wind velocity is measured with a hot-wire anemometer (test 405i). The flow Reynolds number is between $10^5$ and $10^6$.

\subsection{Aerosol measurements}

Aerosols are produced from heated oil vapor injected at a controlled rate and power through a $6\;{\rm mm}$ nozzle. A cloud of droplets nucleates at a few millimeters downstream. The test section is illuminated from above by a LED array with a diffuser. The cloud is made dilute enough so that double scattering is negligible and the light intensity scattered is linear with respect to the local aerosol concentration. High resolution pictures with a $1$ or $2\;{\rm s}$ exposure time are taken with a Digital Single-Lens Reflex camera. In order to keep the image data linear with respect to the light intensity, the raw images files are debayered using a bilinear approximation and no further image processing is performed. We keep only the green channel to measure the light intensity. We average sets of $10$ pictures taken in the same conditions to achieve a satisfying statistical convergence. Residual background light is removed by substracting the image of the tunnel without droplets. The illumination intensity profile along the tunnel axis $I_0(x)$ is calibrated using a diffusive object. $I_0(x)$ is flat, except at the start and the end of the measurement section where it decreases because of greater distance to the illumination source.

\begin{figure}[h!]
\includegraphics[width=7 cm]{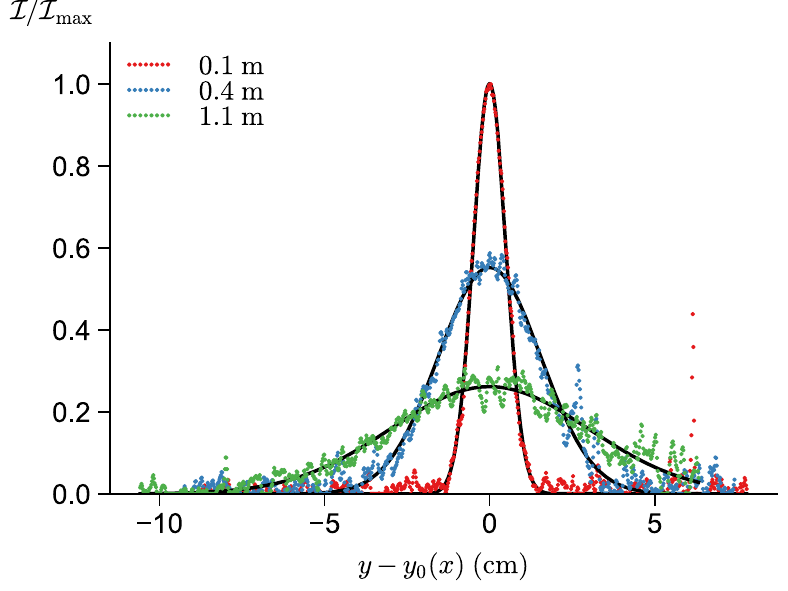}
\caption{Transverse intensity profiles at various distances from the oil outlet. The best fit by a Gaussian gives ${\mathcal C}(x)$, $\sigma_R(x)$ and $y_0(x)$.}
\label{fig:gauss}
\end{figure}

The intensity ${\mathcal I}(x,y)$ of a single pixel is an integral over the $z$ axis of the scattered intensity, which is Gaussian as the concentration is Gaussian along a transverse section. Integrating it over $z$, the image intensity field ${\mathcal I}(x,y)$ is Gaussian along $y$:
\begin{equation}
{\mathcal I}(x,y)=I_0(x) {\mathcal C}(x) \sigma_R(x) \exp{\left(-\frac{\left(y-y_0(x)\right)^2}{2\sigma_R^2(x)}\right)}
\label{eq:gauss}
\end{equation}
As shown on Figure~\ref{fig:gauss}, we fit each transverse pixel line over $y$ by this corrected Gaussian profile to extract ${\mathcal C}(x)$, which is proportional to the concentration on the axis, and the radius $\sigma_R(x)$. ${\mathcal C}(x) \sigma_R^2(x)$ is a constant independent of $x$: conservation of mass is respected by our measurement technique.

\subsection{CO\textsubscript{2} measurements}

Solid CO\textsubscript{2} pellets are sublimated inside a pressurized container heated with a controlled power. A variable area flow meter controls the gas injection rate. The gas is injected at room temperature into the test section. CO\textsubscript{2} is then sampled along the dispersion cone axis by slowly suctioning air into a $60 \; {\rm mL}$
syringe with a long needle. Three replicates are taken at the same point. A laser beam pointing along the flow axis helps position precisely the needle tip. After sampling, syringes are sealed until analysis.

\begin{figure}[h!]
\includegraphics[width=7 cm]{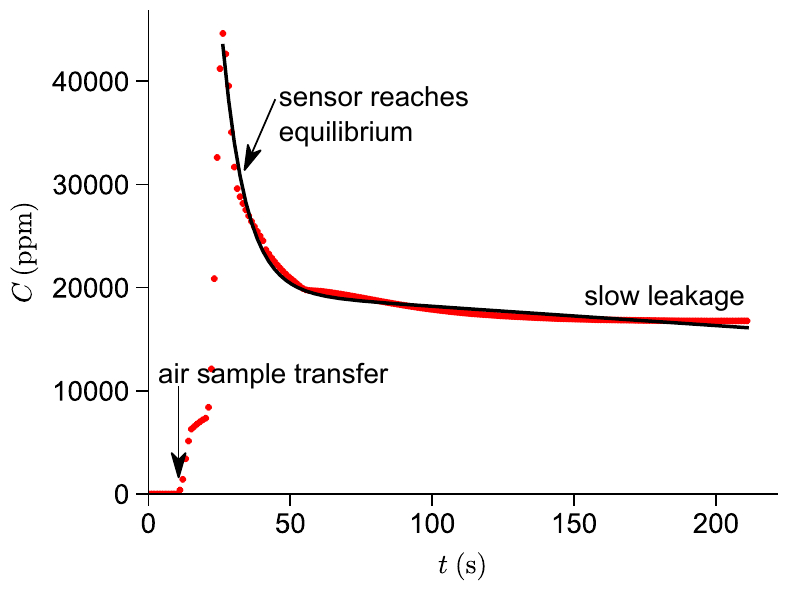}
\caption{Red dots: typical measured CO\textsubscript{2} concentration inside the vacuum chamber. The concentration rises rapidly as CO\textsubscript{2} is introduced and passes by the sensor, then the sensor equilibrates with an exponential decay. There remains a very slow decay, linear in time, due to leakage and adsorption. Solid black line: best fit by an exponential decay superimposed with a linear decrease in time.}
\label{fig:co2concentration}
\end{figure}

The content of the syringe is transferred into a $300 \; {\rm mL}$ chamber equipped with a non-dispersive CO\textsubscript{2} infrared sensor with a range of $0$--$\SI{50000}{ppm}$ and with an absolute pressure sensor. Using a vacuum pump, a primary vacuum is made in the sealed chamber. Then, the $60 \; {\rm mL}$ sample is sucked into the chamber. Finally, nitrogen is immediately added to the chamber to reach the ambient pressure. The CO\textsubscript{2} concentration is recorded as a function of time. After a few minutes, the sensor equilibrates and the concentration reaches a steady state. Figure~\ref{fig:co2concentration} shows a typical temporal evolution of the measured CO\textsubscript{2} concentration. It rises rapidly as the sensor is near the entrance of the chamber, then the concentration homogenizes inside and the sensor relaxes to equilibrium with an exponential decay. There remains a small amount of leakage and adsorption that we fit with a linear decrease in time.

Since we measure its volumetric injection rate $Q$, the absolute volumic concentration of CO\textsubscript{2} is therefore known along the axis. This is not the case for the oil droplet aerosol, where only the relative concentration is known. The measured concentration on the axis $C_M(x)$ can be related to the transverse concentration profile:
\begin{equation}
C(r, x) = C_M(x) e^{-r^2/2\sigma_R^2}
\end{equation}
Conservation of mass imposes that
\begin{equation}
\int_0^\infty 2\pi r \mathrm d r C_M e^{-r^2/2\sigma_R^2} = 2 \pi \sigma_R^2(x) C_M(x)
\end{equation}
is set by the injection rate, i.e.
\begin{equation}
2 \pi \sigma_R^2(x) C_M(x) = \frac{Q}{\overline v}
\end{equation}
We therefore plot $\overline v C_M /Q$ for CO\textsubscript{2} and $1/2\pi \sigma_R^2$ for oil droplets.

\section{Turbulent dispersion: Lagrangian approach}

Turbulent diffusion, like many diffusive processes \citep{risken_fokker-planck_1989}, can be seen through two complementary lenses: a Langevin equation describing particle dynamics and a Fokker-Planck equation describing the spacial evolution of the concentration field. We develop these two approaches here. As turbulence acts as a fluctuating noise, in both cases we use the Reynolds decomposition \citep{lesieur_turbulence_2008} to split any field or dynamical variable $A$ into its ensemble average over many realizations of the flow $\overline A$ and its fluctuating part with zero mean $A'$: $A = \overline A + A'$.

We first consider a single particle of diameter $d$, density $\rho_p$, in a fluid of density $\rho_f$ and viscosity $\eta$. Turbulent diffusion for particles without inertia is a correlated Brownian motion \citep{batchelor_diffusion_1949,batchelor_application_1950,batchelor_diffusion_1952}. The simplest form of a Langevin equation that leads to a finite correlation time gives an exponentially decaying velocity correlation function \citep{risken_fokker-planck_1989,sawford_reynolds_1991,sawford_turbulent_2001,salazar_two-particle_2009}:
\begin{equation}
\label{eq:correlation_velocity}
\overline{\mathbf{v^{\prime}}(t)\mathbf{v^{\prime}}(t+\tau)}=\sigma_V^2\exp(-\tau/{\mathcal T})
\end{equation}

The correlation time ${\mathcal T}$ is called the Lagrangian integral time scale, defined in the general case as:
\begin{equation}
{\mathcal T}=\frac{1}{\sigma_V^2}\;\int_0^\infty \overline{\mathbf{v^{\prime}}(t)\mathbf{v^{\prime}}(t+\tau)} \mathrm d\tau
\end{equation}
The dispersion of fluid particles injected at a source point at time $t=0$ is given by Taylor's theorem \citep{taylor_diffusion_1922,sawford_turbulent_2001,salazar_two-particle_2009}:
\begin{equation}
\frac{d \overline{{\mathbf r}^{2}(t)}}{d t}=2 \overline{{\mathbf r}(t) {\mathbf v}(t)}=2 \int_{0}^{t} \overline{{\mathbf v}\left(t^{\prime}\right) {\mathbf v}(t)} \mathrm d t^{\prime} .
\end{equation}
Applied to equation~(\ref{eq:correlation_velocity}) at time $t=x/\overline v$, the transverse spread is \citep{sawford_turbulent_2001,du_estimation_1995,salazar_two-particle_2009,poydenot_turbulent_2021-1}:
\begin{equation}
\sigma_R^{2}= \frac{2}{3}\sigma_V^2 {\mathcal T}^2 \left[\exp \left(-\frac{x}{{\overline v}{\mathcal T}}\right) +\frac{x}{{\overline v}{\mathcal T}}-1\right]
\label{eq:EqDiffusionRadius}
\end{equation}
At short and long distances compared to $\overline v \mathcal T$, $\sigma_R$ can be rewritten as the mean square displacement at time $t=x/\overline v$ of a diffusive process with an effective diffusion coefficient $D$:
\begin{equation}
\sigma_R^2 \sim D \frac{x}{\overline v}
\end{equation}
The effective diffusion coefficient is constant at long distances $x \gg \overline v \mathcal T$, as expected for an uncorrelated random walk of velocity $\sigma_V$ and mean free path $\sigma_V \mathcal T$:
\begin{equation}
D \sim \sigma_V^2 \mathcal T
\end{equation}
At short distances $x \ll\overline v \mathcal T$, particle motion is correlated so that the mean free path is rather the distance traveled $\sigma_R$:
\begin{equation}
D \sim \sigma_V \sigma_R
\end{equation}
This gives the two scalings reported in the article.

Unlike passive scalars, particles are subjected to gravity and inertia. The equation of motion of a single particle reads \citep{maxey_equation_1983}
\begin{equation}
\label{eq:pdf_particule}
\frac{\mathrm d}{\mathrm d t} \mathbf r=\mathbf v, \quad \frac{\mathrm d}{\mathrm d t} \mathbf v=-\frac{1}{\tau_S}(\mathbf v-\mathbf u[\mathbf r]) + \left(1-\frac{\rho_f}{\rho_p}\right) \mathbf g
\end{equation}
where $\mathbf{u} = \overline{\mathbf{u}}+\mathbf{u'}$ is the fluid velocity and $\tau_S = \rho_p d^2/18 \eta$ the Stokes times, which is the particle response time to a change in the fluid velocity given by the Stokes force $-3\pi \eta d (\mathbf v - \mathbf u)$. $\left(1-\rho_f/\rho_p\right) \mathbf g$ is a buoyancy term. In the absence of turbulence, particles fall with a velocity
\begin{equation}
\label{eq:uchute}
\mathbf v_{\rm fall}=\frac{(\rho_p - \rho_f)\mathbf g d^2}{18\eta}.
\end{equation}
If the fluid presents a constant mean velocity, particles move on average at a velocity $\overline{\mathbf{v}} = \overline{\mathbf{u}}+\mathbf v_{\rm fall}$. We consider again for simplicity that the fluid velocity correlation function (but not the particle velocity correlation function $\overline{\mathbf{v^{\prime}}(t)\mathbf{v^{\prime}}(t+\tau)}$) decays exponentially over a time ${\mathcal T}$:
\begin{equation}
\label{eq:CorrelVelo}
\overline{\mathbf{u^{\prime}}(t)\mathbf{u^{\prime}}(t+\tau)}=\sigma_U^2 \exp(-\tau/{\mathcal T})
\end{equation}

The aerosol phase forms when turbulent velocity fluctuations are large enough to counteract particle settling \citep{friedlander_smoke_2000} i.e. when $\sigma_U>v_{\rm fall}$.

The fluctuating velocity obeys:
\begin{equation}
\mathbf{v'}(t) = \frac{1}{\tau_S}\int_{-\infty}^t \mathbf{u'}(t') \exp{\left(-\frac{t-t'}{\tau_S}\right)} \mathrm d t'
\end{equation}The particle velocity correlation function is the low-passed fluid velocity correlation:

\begin{equation}
\overline{\mathbf{v^{\prime}}(t)\mathbf{v^{\prime}}(t+\tau)} = \frac{1}{2\tau_S} \int_{-\infty}^{\tau} \overline{\mathbf{u'}(t_2)\mathbf{u'}(t_2+t_1)} \exp{\left(-\frac{\tau-t_1}{\tau_S}\right)} \mathrm d t_1
\end{equation}
 Performing the Reynolds decomposition $\mathbf{v}=\overline{\mathbf{v}}+ {\mathbf{v'}}$ and using the Fourier transform of the fluid velocity autocorrelation function, we find:
\begin{equation}
\overline{\mathbf{v^{\prime}}(t)\mathbf{v^{\prime}}(t+\tau)} = \frac{\sigma_U^2 }{1-\frac{\tau_S^2}{{\mathcal T}^2}} \left(\exp{(-\tau/\mathcal T)}-\frac{\tau_S}{{\mathcal T}}\exp{(-\tau/\tau_S)} \right)
\end{equation}
Inertia acts as a low-pass filter of the fluid velocity \citep{bec_turbulent_2010}. The dimensionless ratio $\mathrm{St} = \tau_S/\mathcal T$ is called the Stokes number and characterizes the relative influence of particle inertia and hydrodynamic drag. If the Stokes number $\mathrm{St}$ is much smaller than $1$, particles presents a negligible inertia and the particle velocity correlation function reduces to the fluid velocity correlation function $\overline{\mathbf{v^{\prime}}(t)\mathbf{v^{\prime}}(t+\tau)} = \overline{\mathbf{u^{\prime}}(t)\mathbf{u^{\prime}}(t+\tau)}$. Conversely, in the limit where the Stokes number $\mathrm{St}$ is much larger than $1$, the correlation function decays exponentially as $\sigma_U^2 \mathrm{St} \exp{(-\tau/\tau_S)}$.

\section{Turbulent dispersion: Reynolds-averaged Eulerian approach}

In the continuum approximation, we introduce the concentration $C(\mathbf r, t)=\overline{C}(\mathbf r, t)+C'(\mathbf r, t)$ of particles and its flux $\mathbf{j}$. The conservation equation reads:
\begin{equation}
\frac{\partial C}{\partial t}+\nabla \cdot \mathbf{j}=0
\end{equation}
Transport by the the average flow leads to a flux equal to the concentration times velocity $\mathbf{j}=C \mathbf{v}$. Consider now thermal diffusion induced by random microscopic velocity fluctuations at $\mathbf{v}=\mathbf{0}$. Would the concentration be homogeneous i.e. constant in space, the diffusive flux would vanish. At leading order, random exchanges between neighboring layers of fluid lead to a flux proportional to the gradient of concentration, oriented from high to low concentration. The phenomenological relation is known as Fick's law:
\begin{equation}
\mathbf{j}=-D_m \mathbf{\nabla} C
\end{equation}
and the molecular diffusion coefficient $D_m$ is assumed to be independent of $C$, which must be true in the dilute limit $C \to 0$. Plugging this relation into the continuity equation, we get the linear diffusion equation:
\begin{equation}
\frac{\partial C}{\partial t}=D_m \nabla^2 C
\end{equation}

With turbulence, the relevant transport equation is obtained by Reynolds averaging the concentration conservation equation:
\begin{equation}
\frac{\partial \overline{C}}{\partial t}+\nabla \cdot(\overline{\mathbf{v}} \overline{C})=\nabla \cdot\left(D_m \nabla \overline{C}-\overline{\mathbf{v^{\prime}} C^{\prime}}\right)
\end{equation}
For the reasons invoked for molecular diffusion, the flux $\overline{\mathbf{v^{\prime}} C^{\prime}}$ can be expressed using a gradient diffusion assumption:
\begin{equation}
\overline{\mathbf{v^{\prime}} C^{\prime}}=-D_t \nabla \overline{C}
\end{equation}
where $D_t$ is the turbulent diffusion coefficient, also called eddy diffusivity. The averaged concentration therefore obeys:
\begin{equation}
\frac{\partial \overline{C}}{\partial t}+\nabla \cdot(\overline{\mathbf{v}} \overline{C})=\nabla \cdot\left(D \nabla \overline{C}\right)
\end{equation}
where $D=D_m+D_t$ is the effective diffusion coefficient. In the Lagrangian approach, we have introduced the decorrelation time $\mathcal T$. For the Eulerian framework, we must introduce the inertial length scale $\mathcal L$; above this space length, fluid velocities are uncorrelated. 

For a point source at the origin in an average flow along the $x$ direction in cylindrical $(r, \theta, x)$ coordinates and neglecting longitudinal diffusion, the average concentration $\overline C$ obeys a convection-diffusion equation which reduces in the steady state to:
\begin{equation}\label{eq:convection-diffusion}
\overline v \frac{\partial \overline C}{\partial x} = r^{-1} \frac{\partial}{\partial r}\left(r\;D\;\frac{\partial \overline C}{\partial r}\right)
\end{equation}
This equation has the same structure as the diffusion equation, except that time is replaced by the space coordinate $x$. In the regime described by Taylor, the turbulent diffusion coefficient $D$ does not depend on $r$, the equation admits an exact solution:
\begin{equation}
\overline C=\frac{q_e C_e}{\pi \sigma_R^2 \bar v}\;\exp\left(-\frac{r^2}{2\sigma_R^2}\right)
\end{equation}
where the multiplicative factor is obtained by identifying the mass flow rate across any section to the source injection rate $q_e C_e$. The dispersion radius $\sigma_R$ obeys the equation:
\begin{equation}
\bar v \frac{\mathrm d \sigma_R^2}{\mathrm d x}=2 D
\label{eq:radiusdiffuse}
\end{equation}

We now consider the opposite limit where there is no mean flow at all, but convective plumes creating turbulent mixing. For simplicity, we can assume that dispersion is homogeneous and isotropic and write the diffusion equation in spherical coordinates:
\begin{equation}\label{eq:convection-diffusion2}
\frac{\partial \overline C}{\partial t} = r^{-2} \frac{\partial}{\partial r}\left(r^2\;D\;\frac{\partial \overline C}{\partial r}\right)
\end{equation}
Again, considering a constant source of flow rate $q_e$ and concentration $C_e$, a steady state solution gradually appears, which obeys \citep{cheng_modeling_2011}:
\begin{equation}
\frac{\mathrm d \overline C}{\mathrm d r}=-\frac{q_e C_e}{4 \pi r^2\;D}
\end{equation}
At large scale, $D\sim\sigma_V \mathcal L$ can be considered as constant so that $\overline C$ decreases as $r^{-1}$. In the intermediate range of scales, using again the ballistic approximation $D\sim \sigma_V r$, $\overline C$ decreases as $r^{-2}$. The scaling laws derived before therefore still hold, but up to a geometrically determined multiplicative constant $\alpha$.

\section{Droplet production}
Coughing, sneezing, singing, speaking, laughing or breathing produce droplets of mucosalivary fluid in two range of sizes. Droplets above $100\;{\rm \mu m}$ are produced by fragmentation of a liquid sheet formed at the upper end of the respiratory tract (i.e. for sneeze) or from filaments between the lips (i.e. plosives consonants when speaking \citep{abkarian_speech_2020,abkarian_stretching_2020}), with an average around $500\;{\rm \mu m}$ \citep{morawska_size_2009,johnson_modality_2011,bourouiba_fluid_2021}. In the first case, the initial sheet is stretched and get pierced. The liquid accumulates by capillarity retraction in a rim, which destabilizes into ligaments \citep{kooij_what_2018,bourouiba_violent_2014}. The latter form droplets by a capillary instability referred to as the beads-on-a-string \citep{scharfman_visualization_2016}. In the second case, a film forms between the lips, which destabilizes into filaments, themselves exhibiting the beads-on-a-string instability.

Droplets below $20\;{\rm \mu m}$ form either by bubble bursting events in the lungs alveoli or by turbulent destabilization of liquid films covering the lower and upper airways. Between $20\;{\rm \mu m}$ and $100\;{\rm \mu m}$, almost no droplets form, indicating well separated mechanisms. For each class of droplets, the distribution around the average has been fitted either by a log-normal distribution, following the idea of a break-up cascade, or by a Gamma distribution, based on the idea that the blobs that make up a ligament exhibit an aggregation process before breaking up \citep{villermaux_fragmentation_2007}. The average droplet size, around $4\;{\rm \mu m}$ results from an interplay between the fluid film thickness, the turbulent stress and the surface tension. Further fragmentation of these droplets can occur in the tract constrictions where air flows at large velocity. Pulmonary surfactant helps reduce the droplet size, and contributes to prevent accumulation of fluid in airways. The main entry zone is through the nasal epithelium for viral strains before Omicron and more specifically a subset of cells of the nasal epithelium expressing both the ACE2 receptor and the TMPRSS2 protease, and the throat epithelium for Omicron, due to a weaker dependency to the TMPRSS2 protease. Other entry zones exist as well as different receptors and proteases \citep{singh_single-cell_2020}. In first approximation, they can be considered as minor routes in the dissemination of the epidemic and in the risk of infection. The emission of mucus droplets containing viral particles in the nasal cavity is dominant, but has not been much investigated so far \citep{li_detecting_2021,ma_coronavirus_2021}.
\begin{figure}[t!]
\includegraphics[width=7 cm]{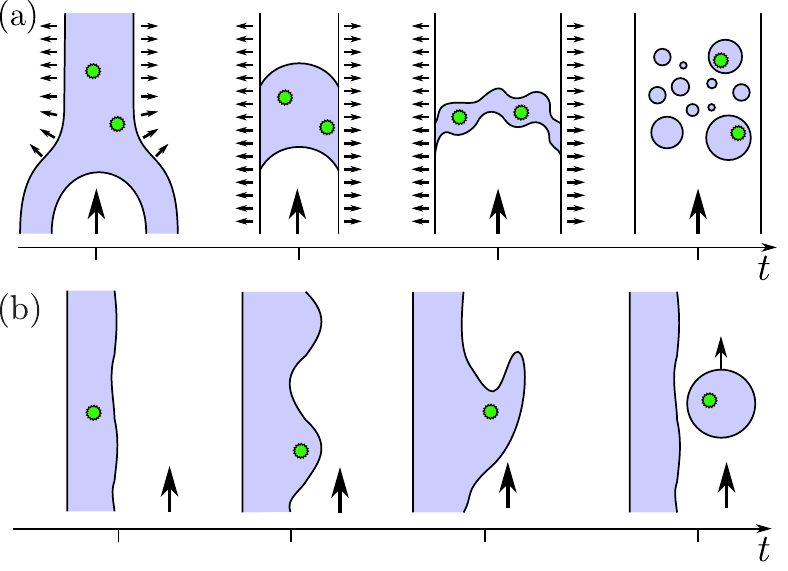}
\caption{(a) Sketch of the aerosol generation mechanism inside the bronchioles. The walls are elastic and collapse during exhalation; when air flows back into the bronchioles, it spreads them apart, leading to the formation of a fluid film which then bursts into submicronic droplets \citep{malashenko_propagation_2009-1,johnson_mechanism_2009,mittal_flow_2020,halpern_nonlinear_2003,haslbeck_submicron_2010}. (b) Sketch of the aerosol generation mechanism inside the upper respiratory tract. Shear destabilizes the respiratory fluid layer lining the walls \citep{kataoka_generation_1983,moriarty_flow-induced_1999,mittal_flow_2020}, and droplets detach from the waves. In both cases, if virions are present, they get entrained when the film forms and are found in the subsequent aerosol droplets.}
\label{fig:SketchAerosol}
\end{figure}

\subsection{Evaporation process and virus inactivation}
We discuss here a basic model of evaporation of droplets, before efflorescence, in humid air. The evaporation of liquid droplets in the air is controlled by the ambient relative humidity $\mathrm{RH}$. Mass transport of water molecules from the droplets to the surrounding air is diffusive; as the drops evaporate, they release latent evaporation heat, which is also conducted away. This cools down the droplets, which in turns lowers the saturation pressure in the immediate surrounding of the drop, slowing down evaporation. Due to this coupled transport, a drop of initial radius $a_0$ shrinks to a radius $a(t)$ as \citep{pruppacher_microphysics_2010,smith_aerosol_2020}
\begin{equation}
 a^2(t) = a_0^2 - 2 D_\mathrm{eff} \left(1-\mathrm{RH}\right) t
\end{equation}
where $D_\mathrm{eff} = 1.3\,10^{-10}\;{\rm m^2/s}$ is an effective diffusion coefficient taking into account both diffusive transport of mass and its slowdown due to evaporative cooling. Evaporation takes places at the surface of the drop, which explains the linear behavior of $a^2$ with $t$. Since the drops are small, evaporation is very fast compared to the time they spend aloft: a $\SI{4}{\mu m}$ drop at $\SI{70}{\%}$ RH completely evaporates in $\SI{0.2}{s}$, while it takes $\SI{20}{min}$ to fall a distance of $\SI{2}{m}$ under its own weight.

The classical picture \citep{wells_air-borne_1934} considers evaporating droplets as independent. This is true for small droplets dispersed inside a room, but not of droplets inside a cough or sneeze spray. In that case, $\mathrm{RH}$ is roughly uniform and close to $\SI{100}{\%}$ inside the aerosol jet, meaning that no evaporation takes places except at the spray boundaries \citep{villermaux_fine_2017,ng_growth_2021}. This makes these drops extremely long-lived, up to a hundred times the isolated drop lifetime \citep{chong_extended_2021,de_oliveira_evolution_2021,smith_aerosol_2020}.

However, virus-laden respiratory droplets do not vanish as they contain viral particles and are not composed of pure water. The mucosalivary fluid is a dilute solution of surfactants, proteins and electrolytes, initially composed of $\sim \SI{99}{\%}$ water in volume. The solutes stabilize drops at a finite radius $a_\mathrm{eq}$, at which they still contain water \citep{mikhailov_interaction_2004}:
\begin{equation}
 a_\mathrm{eq} = \left(\frac{M_w}{\rho_w} \sum_i \frac{\nu_i c_i}{M_i}\right)^{1/3}\; \frac{a_0}{\left(1-\mathrm{RH}\right)^{1/3}} 
\end{equation}
The sum is done over all solutes $i$. $c_i$ is the mass concentration of solute $i$, $M_i$ its molar mass, $\nu_i$ its degree of dissociation ($2$ for NaCl). We model respiratory fluid by a mixture of NaCl and the total average protein content \citep{nicas_toward_2005,vejerano_physico-chemical_2018}: $c_\text{electrolytes} = \SI{9}{g/L}$ (physiological NaCl concentration), $c_\text{proteins} = \SI{70}{g/L}$, $M_\text{proteins} = \SI{70}{kg/mol}$. This gives $ a_\mathrm{eq} \approx 10^{-1} a_0\;\left(1-\mathrm{RH}\right)^{-1/3}$: typically, at medium relative humidity, aerosol drops formed at $5\;{\rm \mu m}$ remain at $1\;{\rm \mu m}$, which is significantly larger than the virus itself. By stabilizing the droplet at a finite radius, solutes reduce the evaporation time to
\begin{equation}
 t_\mathrm{ev} = \frac{a_0^2}{2 D_\mathrm{eff} (1-\mathrm{RH})}\left(1-\left(\frac{a_\mathrm{eq}}{a_0}\right)^2\right)
\end{equation}
The solute effect on the evaporation time is small at low $\mathrm{RH}$ since the drop shrinks by a lot; at $99\;\%$ RH, a $\SI{4}{\mu m}$ drop has its evaporation time increased by $80\;\%$.

In aerosol droplets at equilibrium, virions are gradually degraded by the damage done by antiviral proteins in saliva \citep{malamud_antiviral_2011} and by UV-B radiation in sunlight or by UV-C. The inactivation rate of enveloped, airborne viruses increases \citep{yang_dynamics_2011,tang_effect_2009,chan_effects_2011,smither_experimental_2020} with $\mathrm{RH}$: this suggests that virions can associate with proteins which protects them both from dessication and antivirals \citep{vejerano_physico-chemical_2018}. Another possible effect of ambient conditions results the intrinsic temperature preference of the coronavirus spike protein \citep{laporte_sars-cov-2_2021}, as the temperature in the nose depends on both temperature and humidity.

\section{Masks and respirators}
\begin{figure}[h]
 \centering
 \includegraphics[width=70mm]{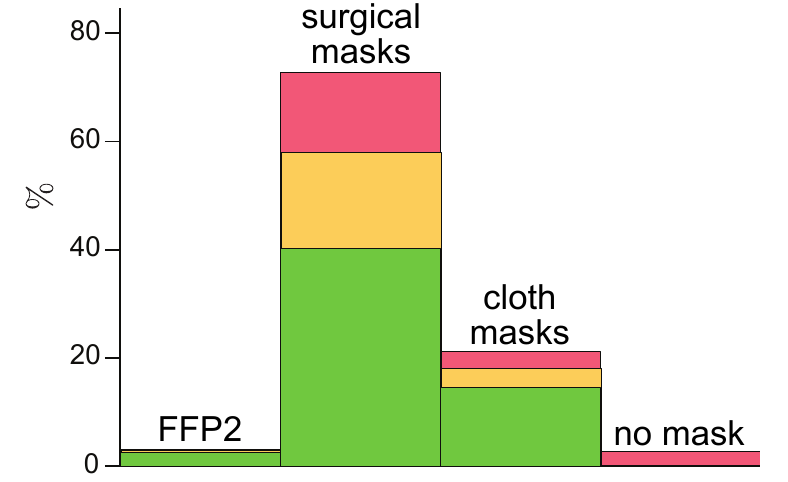}
 \caption{Mask wearing in public spaces of the Paris suburban region (Ile-de-France) where face masks are mandatory, mid April 2021. $N=1708$. Green: correct fitting. Orange: centimeter-scale leakage along the nose or the cheeks. Red: No fitting at all.}
 \label{fig:maskpdf}
 \end{figure}

$\lambda_i$ and $\lambda_e$ are respectively the inhalation and exhalation filtration factor of the face coverings worn by the two people. Filtration factors are determined by the material properties, the fitness of the face coverings and the respiratory activity at play. Cloth masks have a material filtration efficiency of $20\%$--$40\%$ \citep{pan_inward_2021,fischer_low-cost_2020,hill_testing_2020,zangmeister_filtration_2020,rengasamy_simple_2010,shakya_evaluating_2017,drewnick_aerosol_2021}, tightly woven material being more efficient. Surgical masks have a $50\%$--$95\%$ material filtration efficiency \citep{hill_testing_2020,lindsley_comparison_2021,pan_inward_2021,oberg_surgical_2008}, and N95/FFP2 respirators $95\%$--$99\%$ \citep{hill_testing_2020,zangmeister_filtration_2020,qian_performance_1998,rengasamy_simple_2010}. Inhalation (i.e. exposure reduction) and exhalation (i.e. aerosol source control) are not symmetrical as they induce different airflows around the face covering. Moreover, the flow is different for instance for coughs and tidal breathing, which is associated with a $20$ times smaller typical flow rate \citep{pohlker_respiratory_2021}. Surgical masks block $56\%$--$86\%$ of cough aerosols and $42\%$--$91\%$ of exhaled aerosols \citep{blachere_face_2021,lindsley_comparison_2021,lindsley_efficacy_2021,asadi_efficacy_2020}; N95 respirators block $73\%$--$99\%$ of cough aerosols \citep{lindsley_comparison_2021,lindsley_efficacy_2021,asadi_efficacy_2020} and $95\%$--$99\%$ of exhaled aerosols \citep{lindsley_comparison_2021,asadi_efficacy_2020}. Efficiencies for exposure reduction are roughly similar at the most penetrating particle size in well-mixed conditions \citep{pan_inward_2021,lindsley_efficacy_2021-2}. Filtration efficiency strongly depends on proper mask fit: cloth and surgical masks tend to be much looser fitted than respirators, thereby greatly reducing their filtration factor compared to their material efficiency \citep{cappa_expiratory_2021}. Improved surgical mask fit by double masking or knotting and tucking can bring their filtration efficiency up to $60\%$--$99\%$ \citep{sickbert-bennett_fitted_2021,brooks_maximizing_2021,clapp_evaluation_2021,blachere_face_2021}. We take as an average $\lambda = \sqrt{\lambda_i \lambda_e}$ for both inhalation and exhalation. $\lambda = 1$ when no mask is worn; $\lambda = 0.70$ for cloth masks ($30\%$ efficiency); $\lambda = 0.28$ for well-fitted surgical masks ($72\%$ efficiency); $\lambda = 0.10$ for N95 respirators ($90\%$ efficiency). For respirators, fit checks are needed to reach high efficiencies, protective enough for workplace aerosol exposure. However, most untrained users achieve on their own $90\%$ and higher filtration, making respirators a valuable choice for the general public \citep{rembialkowski_impact_2017,brosseau_fit_2010}.

We have performed a quantitative study of face mask wearing in public spaces around Paris. Investigators pair up in a public place and observe the people around them without taking notes, keeping their count. One of the two investigators is familiar with the public place and the other not. They decide on the type of mask they wear and the most appropriate memorization technique, after a reconnaissance visit. Every 10 people, they discreetly write down the quantitative measurements on the type of mask they wear and on the mask fitting. Every twenty minutes, in a calm environment, without witnesses, they discuss the social characteristics of the place under study, at this time of day and this day of the week, in order to prevent perception biases. Ideally, the investigators fill, for each person, the following categories: gender, age, occupational characteristics, visible religious attributes, spoken language, alone or accompanied, behavior. Special attention is paid to the difference between externally imposed rules vs internalized rules, when noting the behaviors in the presence or not of a control \citep{goffman_les_1988,hall_hidden_1990,winkin_nouvelle_2000}.

\begin{figure}[h]
 \centering
 \includegraphics[width=70mm]{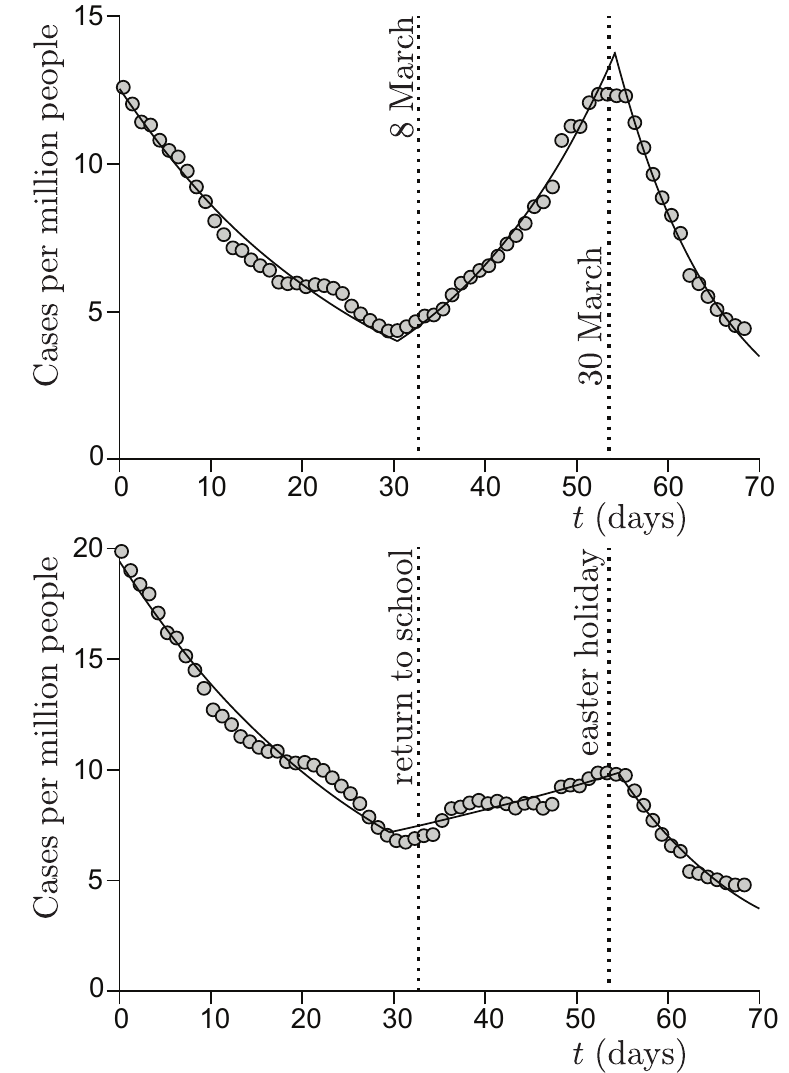}
 \caption{Cases per million people in the United Kingdom, from 1 February to 12 April 2021. Top panel: schoolchildren aged 10-14, with no mandatory masks. The best fit by an exponential provides the reproduction number: $R= 1.34\pm0.04$ during the school period vs $R= 0.81\pm0.03$ before and $R= 0.70\pm0.03$ after. Bottom panel: schoolchildren aged 15-19, with mandatory masks. The best fit by an exponential provides the reproduction number: $R= 1.07\pm0.04$ during the school period vs $R= 0.82\pm0.03$ before and $R= 0.70\pm0.03$ after.}
 \label{fig:UKschools}
 \end{figure}

Social aspects will be discussed in a separate paper. Here, it is used to ensure that face mask wearing statistics reflect social activities at different times of the week, in the Paris region. Figure~\ref{fig:maskpdf} shows the resulting histogram. Only indoor public spaces where masks are mandatory have been included. FFP2 respirators represent $3\;\%$ only of face barriers and correctly worn protective coverings (surgical or FFP2) only $42\;\%$. We can estimate that the mean filtration factor $\lambda_i \lambda_e$ is around $0.25$ only, which gives a dose reduction by a factor of $4$ in French public spaces where masks are mandatory. As even minute differences in respirator fit can lead to large variations in filtration, this visual method of estimating mask fit could lead to overestimating the average filtration efficiency. In Georgia, USA during November-December 2020, mask wearing for students and staff members was associated with a $2.7$-fold reduction in COVID-19 cases \citep{gettings_mask_2021}. In the UK, between February and April 2021, schoolchildren aged 10-14 did not wear masks while those aged 15-19 did. Figure \ref{fig:UKschools} compares the cases in these two populations. The reproduction rate that can be attributed to schools is roughly halved by masks, which gives a mean filtration factor $\lambda^2 = 0.5$. We have observed that young people poorly fit their masks, which could explain the discrepancy with the expected filtration factor. Another explanation could be that an important part of the transmission takes places during unmasked lunch time.

 \section{Risk reduction techniques}
 \begin{figure*}[bt]
 \centering
 \includegraphics[width=14cm]{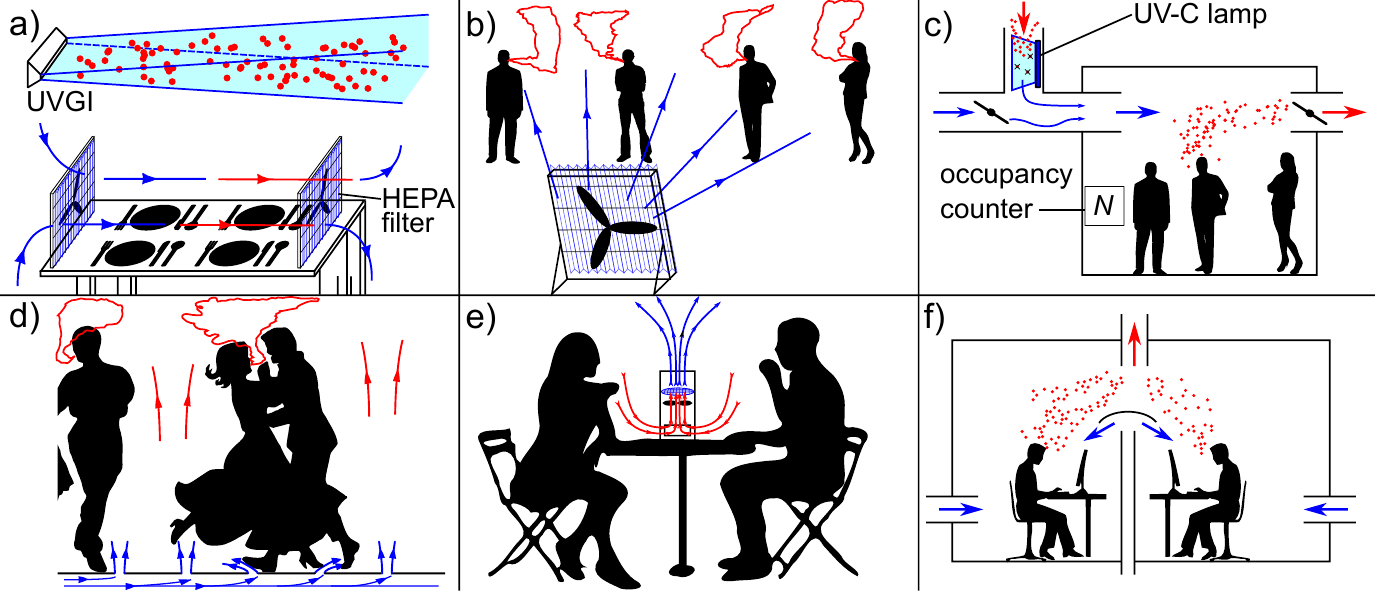}
 \caption{Hydrodynamic solutions to mitigate short-range transmission by dispersing aerosols away from individuals. (a) Fans force the circulation of air through HEPA filters at a cafeteria table, with an Upper-Room Ultraviolet Germicidal Irradiation (UVGI). (b) A large fan with a filter is slightly tilted upward and aimed at a line of static people waiting. The cone of aerosols they emit cannot reach anyone in the line. (c) Ventilation controlled by the occupancy number, to reduce energy consumption. Recycled air inside ventilation ducts is decontaminated by UV-C light. (d) An air flow inside the floor of a club pushes air up through small holes, preventing aerosols from spreading laterally and dispersing them towards the ceiling where the ventilation system can remove them. (e) A fan at a caf\'e table pulls air, filters it and expels it upwards. (f) Fresh air is injected at the bottom of the room and supplied where needed to create air flows that protect each occupant of the room individually \citep{morawska_paradigm_2021}.}
 \label{fig:mesuressecurisation}
 \end{figure*}
 
\begin{figure}[h]
 \centering
 \includegraphics[width=80mm]{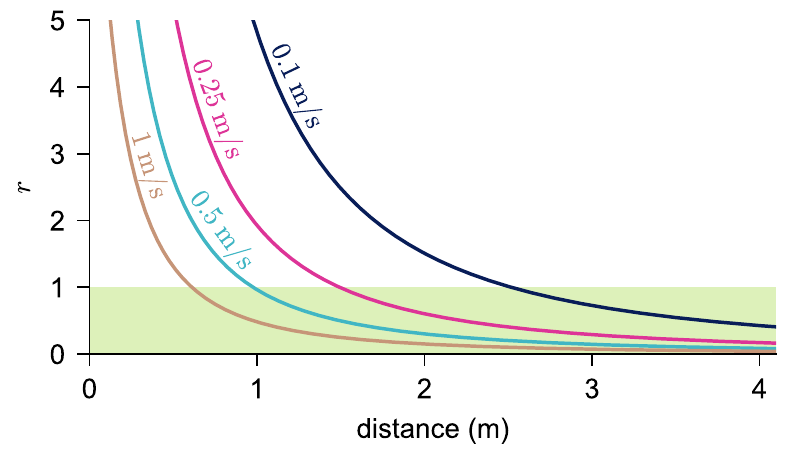}
 \caption{Transmission risk $r$ for Omicron BA.1 ($\bar h = 2800$) as a function of the distance between people inside the public space, for the observed mask wearing shown in Figure \ref{fig:maskpdf} ($\lambda^2 = 0.2$) and different wind speeds. Green area: acceptable risk, ie. $r < 1$.}
 \label{fig:risk_distance}
 \end{figure}

The transmission risk outdoors, without masks, is real at short distances and for long periods of time. Figure \ref{fig:risk_distance} shows the risk as a function of distance. The risk is above $1$ at low wind speeds below $2.5\;{\rm m}$. Static crowds without masks must therefore be avoided. It is important for people to learn how to take into account the wind strength and direction for static outdoors activities, in particular if they sing, eat, or drink for a long duration. As the risk outdoors is entirely at short range, large fans may be used to reduce the risk of a bar terrace or static queue in front of a shop (figure~\ref{fig:mesuressecurisation} panel b). The more these fans induce turbulent fluctuations, rather than an average flow, the better they are. They must be oriented upwards to change the wake direction. For outdoor dance floors, injection of air at high flow rate, say, $1-10\;{\rm m^3/hour/person}$ may be sufficient to reduce the risk.

The absence of masks when eating and drinking poses a specific problem of aerosol risk reduction. In particular, collective catering facilities are amongst the most important places of high transmission risk. It is possible to use HEPA filtered air purifiers arranged to provide air free of viral particles and suck out stale air \citep{morawska_paradigm_2021} (Figure~\ref{fig:mesuressecurisation} a, e).

Disposable respiratory personal protection equipment is expensive, but various protocols involving UV-C, heat and hydrogen peroxide vapor have been designed to extend its lifetime cycle and use, and could be considered for the general public \citep{cdc_decontamination_2020,john_scalable_2021}.

\bibliography{ShortLongRangeCovid_nodoi}

\end{document}